\documentclass{article}

\usepackage{microtype}
\usepackage{graphicx}
\usepackage{multirow}
\usepackage{xcolor}
\usepackage{subfig}

\usepackage{booktabs} 
\usepackage[most]{tcolorbox}

\usepackage{hyperref}

\usepackage[accepted]{icml2024}

\usepackage{amsmath}
\usepackage{amssymb}
\usepackage{mathtools}
\usepackage{amsthm}

\usepackage[capitalize,noabbrev]{cleveref}

\theoremstyle{plain}
\newtheorem{theorem}{Theorem}[section]
\newtheorem{proposition}[theorem]{Proposition}

\theoremstyle{definition}
\newtheorem{definition}[theorem]{Definition}

\theoremstyle{remark}

\usepackage[textsize=tiny]{todonotes}

\newcommand{\Ham}{\hat{\mathcal{H}}}
\newcommand{\Obs}{\hat{\mathcal{O}}}
\newcommand{\Graph}{{\mathcal{G}}}
\newcommand{\edges}{{\mathcal{E}}}
\newcommand{\vertices}{{\mathcal{V}}}
\newcommand{\e}{\mathbf{\text{e}}}
\newcommand{\Sum}{\displaystyle\sum}
\newcommand{\Prod}{\displaystyle\prod}

\newcommand{\Sin}{\sin\theta}
\newcommand{\Cos}{\cos\theta}
\newcommand{\Tan}{\tan\theta}

\newcommand{\bra}[1]{\left\langle #1\right|}
\newcommand{\ket}[1]{\left|#1\right\rangle}

\newcommand{\ii}{\textbf{i}}

\newcommand{\first}[1]{\textcolor{blue}{#1}}
\newcommand{\second}[1]{\textcolor{red}{#1}}
\newcommand{\third}[1]{\textcolor{orange}{#1}}

\def\<{\langle}
\def\>{\rangle}

\newcommand{\be}{\begin{equation}}
\newcommand{\ee}{\end{equation}}
\newcommand{\bea}{\begin{eqnarray}}
\newcommand{\eea}{\end{eqnarray}}

\icmltitlerunning{Quantum Positional Encodings for Graph Neural Networks}

\begin{document}

\twocolumn[
\icmltitle{Quantum Positional Encodings for Graph Neural Networks}




\begin{icmlauthorlist}
\icmlauthor{Slimane Thabet}{pasqal,lip6}
\icmlauthor{Mehdi Djellabi}{pasqal}
\icmlauthor{Igor Sokolov}{pasqal}
\icmlauthor{Sachin Kasture}{pasqal}
\icmlauthor{Louis-Paul Henry}{pasqal}
\icmlauthor{Loïc Henriet}{pasqal}
\end{icmlauthorlist}

\icmlaffiliation{pasqal}{Pasqal}
\icmlaffiliation{lip6}{Sorbonne University}

\icmlcorrespondingauthor{Slimane Thabet}{slimane.thabet@pasqal.com}

\icmlkeywords{Machine Learning, ICML}

\vskip 0.3in
]



\printAffiliationsAndNotice{\icmlEqualContribution} 


\begin{abstract}
In this work, we propose novel families of positional encodings tailored to graph neural networks obtained with quantum computers. 
These encodings leverage the long-range correlations inherent in quantum systems that arise from mapping the topology of a graph onto interactions between qubits in a quantum computer. 
Our inspiration stems from the recent advancements in quantum processing units, which offer computational capabilities beyond the reach of classical hardware. 
We prove that some of these quantum features are theoretically more expressive for certain graphs than the commonly used relative random walk probabilities. 
Empirically, we show that the performance of state-of-the-art models can be improved on standard benchmarks and large-scale datasets by computing tractable versions of quantum features. 
Our findings highlight the potential of leveraging quantum computing capabilities to enhance the performance of transformers in handling graph data.
\end{abstract}

\section{Introduction}
Graph machine learning (GML) is an expanding field of research with applications in chemistry \citep{mpnn}, biology \citep{zitnik2018modeling}, drug design \citep{konaklieva2014molecular}, social networks \citep{scott2011social}, computer vision \citep{harchaoui2007image} and science \citep{sanchez2020learning, xu2018powerful}. 
In the past few years, significant effort has been put into the design of Graph Neural Networks (GNNs)~\citep{bookgnn}. 
The objective is to learn suitable representations that enable efficient solutions to the original problem.

Message Passing Neural Networks (MPNN) \citep{gcn, graphsage, gat, mpnn} is the first and most common approach to build GNNs.
This method exhibits several recognized limitations \citep{zhu2020beyond, chen2020measuring, topping2021understanding},  
that the research community is actively exploring in order to find solutions. 
The key idea is to expand aggregation beyond neighbouring nodes by incorporating information related to the entire graph or a more extensive portion of it. 
Graph Transformers were created according to these requirements, showing success on standard benchmarks \citep{ying2021transformers, rampavsek2022recipe}. 
Among the myriad of proposed architectures, the Graph Inductive Bias Transformer (GRIT)~\citep{ma2023graph} stands out for its impressive generalization capacity.
Similarly to their counterparts in natural language processing, graph transformers compute positional encodings (PE) that are concatenated to node features. 
The most commonly used PEs are the eigenvectors of the laplacian matrix \cite{rampavsek2022recipe, kreuzer2021rethinking} and the random walk probabilities \cite{ma2023graph, graph_mlp_mixer, DwivediL0BB22}.



The goal of this work is to leverage new types of structural features as positional encodings emerging from quantum physics that can be obtained using quantum computers. 
The rapid development of quantum computers during the previous years indeed provides the opportunity to compute features that would be otherwise intractable. 
These features contain complex topological characteristics of the graph, and their inclusion has the potential to enhance the model's quality, reduce training or inference time, and decrease energy consumption. 
The idea of using quantum states containing topological features about the graphs has already been explored theoretically in \cite{Verdon19, Schuld20, Henry21, thabet2022extending}, and experimentally implemented in \cite{albrecht2023quantum}. 
In this work, we propose a method similar to \cite{thabet2022extending} that circumvents the difficulties in training quantum circuits (e.g., expensive computation of quantum gradients, noise-induced barren plateaus, only to name a few~\cite{Cerezo2022}).
In addition, we prove the superiority of our method on strongly regular graphs (SRGs) and perform benchmarks on large-scale datasets.

The paper is organized as follows: In Sec.~\ref{sec:related_work}, we provide a concise overview of the existing research on graph transformers, along with references to the latest developments in quantum graph machine learning, and position our contributions in this landscape.
Sec.~\ref{sec:method_theory} describes the core methodology that we propose.
It details ways to construct a quantum state from a graph and explains how positional encodings can be extracted from that quantum state.
In Sec.~\ref{theory_section}, we detail the properties of quantum positional encodings and show that they are superior to most other classical methods on SRGs. 
Finally, Sec.~\ref{sec:expe} presents the outcomes of our numerical experiments and includes discussions of the results. The code to run all the experiments is available in the supplementary materials.

\begin{figure*}[ht!]
\centering
\includegraphics[width=0.85\textwidth]{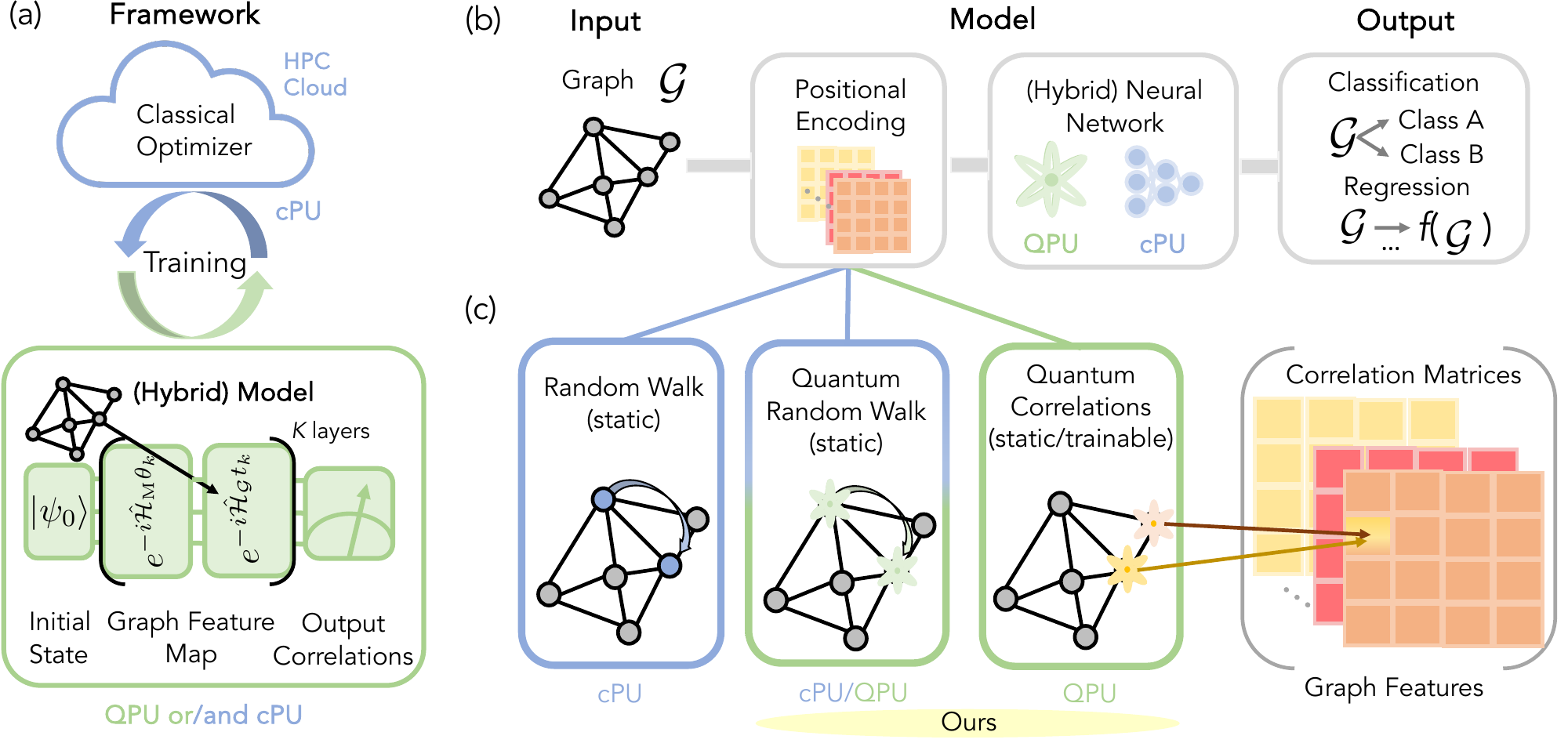}
\caption{
Summary of our method. 
\textbf{(a)} Our hybrid quantum-classical framework utilizes a classical computer for parameter optimization (if required) and employs a hybrid model using a Quantum Processing Unit (QPU) and a CPU and/or GPU, denoted as classical Processing Unit (cPU). In our quantum graph NN, we initialize QPU at a quantum state $|\psi_0\rangle$, apply a mixing Hamiltonian $\Ham_{\rm M}$ evolution for a duration $\theta$, and utilize a Hamiltonian $\Ham_\Graph$ evolution for the graph feature map with a duration $t$. 
$K$ layers are used to obtain a sufficiently expressive quantum model.
Finally, the output is obtained by measuring correlators, e.g., $\langle Z_i Z_j\rangle$.  
See Section \ref{sec:QGML} for details.
\textbf{(b)} Static or trainable PE is constructed for a graph $\Graph$ via \textbf{(c)} (quantum) random walk (static PE) or a quantum graph NN (static/trainable PE), which computes quantum correlations. Note that our PEs are not restricted to classical models (such as the transformer studied in this work) but are also applicable to all quantum models.
}
\label{fig:schema}
\end{figure*}

\section{Related works and contributions}
\label{sec:related_work}

\subsection{Graph Transformers}
Several efforts have been made by the machine learning community to go beyond MPNNs due to several issues including the theoretical limitations of expressivity by the Weisfeiler-Lehman (WL) \citep{morris2019weisfeiler}
test, oversmoothing \citep{chen2020measuring}, oversquashing \citep{topping2021understanding}, and difficutlties on heterophilic data \citep{zhu2020beyond}.
Inspired by the success of transformers in natural language processing \citep{vaswani2017attention, alayrac2022flamingo}, new architectures of GNNs have been proposed to allow an all-to-all aggregation between the nodes of the graphs, the so-called graph transformers (GT) \citep{dwivedi2020generalization, dwivedi2021graph, rampavsek2022recipe, kreuzer2021rethinking, zhang2023rethinking, ma2023graph}.
However, due to the quadratic cost of computing the attention process, they are not applicable to large-scale graphs of millions of nodes and more. 
It has been shown that GTs that include graph inductive biases such as MP modules perform better than those that do not \citep{rampavsek2022recipe, ma2023graph}.

\subsection{Positional and Structural Encoding}
Positional or structural embeddings are features computed from the graph that are concatenated to original node or edge features to enrich GNN architectures (either MPNN or GT). 
These two terms are used interchangeably in the literature and we denote them as "positional encodings"  (PEs) in the rest of this work. 
PEs can include random walk probabilities \citep{rampavsek2022recipe, ma2023graph}, spectral information \citep{dwivedi2020benchmarking, rampavsek2022recipe, kreuzer2021rethinking}, shortest path distances \citep{li2018deeper}, or heat kernels \citep{mialon2021graphit}. 
They can also be learned \citep{dwivedi2021graph}. 
We detail below the most common PEs used in the literature.  \\

\textbf{Laplacian Eigenvectors.}
 The spectral information of the graph can be used as PE, more precisely the eigenvectors of the Laplacian matrix with the smallest eigenvalues, or laplacian eigenvectors (LE). 
For a line graph, the laplacian eigenvectors almost correspond to positional embeddings in the transformer architecture for sequences \citep{vaswani2017attention}. 
 The main issue of this encoding is to ensure that the model remains invariant by changing the sign of eigenvectors, and a solution has been proposed by \citep{lim2022sign}.\\

\textbf{Relative Random Walk Probabilities (RRWP).} 
The authors of \citep{ma2023graph} introduced the RRWP with which they initialize their model. For a graph $\mathcal{G}$, let $A$ be the adjacency matrix and $D$ the degree matrix. Let $P$ be a 3 dimensional tensor such that $P_{k, i, j} = (M^k)_{ij}$ with $M=D^{-1}A$. For each pair of node $(i, j)$, we associate the vector $P_{:, i, j}$, i.e., the concatenation of the probabilities for all $k$ to get from node $i$ to node $j$ in $k$ steps in a random walk. $P_{:, i, i}$ is the same as the Random Walk Structural Encodings (RWSE) defined in \citep{rampavsek2022recipe}. 
The authors of \citep{ma2023graph} highlight the benefits of RRWP. 
They prove that the Generalized Distance WL (GD-WL) test introduced by \citep{zhang2023rethinking} with RRWP is strictly more powerful than GD-WL test with the shortest path distance, and they prove universal approximation results of multi-layer perceptrons (MLP) initialized with RRWP. They also achieve state of the art results on most of benchmark datasets.

\subsection{Quantum Computing for Graph Machine Learning}
In recent years, quantum machine learning has seen a fast development with both theoretical and experimental advances\,\citep{Huang22,Cerezo2022}.
Using quantum computing for machine learning on graphs has already been proposed in several works, as reviewed in \,\citep{Tang22}.
The authors of \citep{Verdon19} realized learning tasks by using a parameterized quantum circuit depending on a Hamiltonian whose interactions share the topology of an input graph.
Comparable ideas were used to build graph kernels from the output of quantum procedures, for photonic \citep{Schuld20} as well as neutral atom quantum processors \citep{Henry21}.
The latter was successfully implemented on quantum hardware~\citep{albrecht2023quantum}.
The architectures proposed in these papers are entirely quantum and only rely on classical computing for the optimization of variational parameters. 
Furthermore, these approaches can only be applied to graph-level tasks, whereas many applications are node-level or edge-level.

\cite{thabet2022extending} proposed a new hybrid architecture that uses the correlation matrix of quantum dynamics containing the information about the graph in the aggregation phase of a larger, entirely classical GNN architecture.
Such a hybrid model presents the advantage of gaining access to hard-to-compute graph topological features through quantum dynamics while benefiting from the power of well-known existing classical architectures. Furthermore, the method presents the advantage to be applicable to all types of tasks, graph-level, node-level or edge-level. 
However, this approach presents some drawbacks. First, the model is difficult to train in simulation because the optimizer of quantum parameters needs to be adjusted separately from the optimizer of classical parameters. 
Secondly, it has been shown that methods to compute the gradients on a quantum computer are hardly scalable.~\cite{abbas2023on}.
Moreover, there were little theoretical indications about the power of the architecture compared to classical methods. 
The experiments have only been made on small datasets compared to the biggest ones available (a few 1000s of graphs and maximum 20 nodes)

\subsection{Contributions}

In light of the related works described above, we propose a method similar to \cite{thabet2022extending} that combines correlations from quantum dynamics and classical GNNs. 
Instead of parameterizing the quantum dynamics and training it as in \cite{thabet2022extending}, we either use random parameterized states or well-chosen quantum states (such as ground states) and use the correlations as positional encoding for graph transformers. 
This approach removes the difficulty of training the quantum states, and can be included in most of classical architectures, which facilitates benchmarking. 
It keeps all the advantages of combining quantum and classical methods while being applicable to any graph machine learning tasks.
In this work, we also give detailed theoretical properties of our positional encoding and show that they are strictly more powerful on some instances of graphs than random walks and spectral methods. 
Finally, we provide benchmarks on large-scale datasets ($\sim$ 3 millions graphs).

\begin{figure}
\centering
\includegraphics[width=\columnwidth]{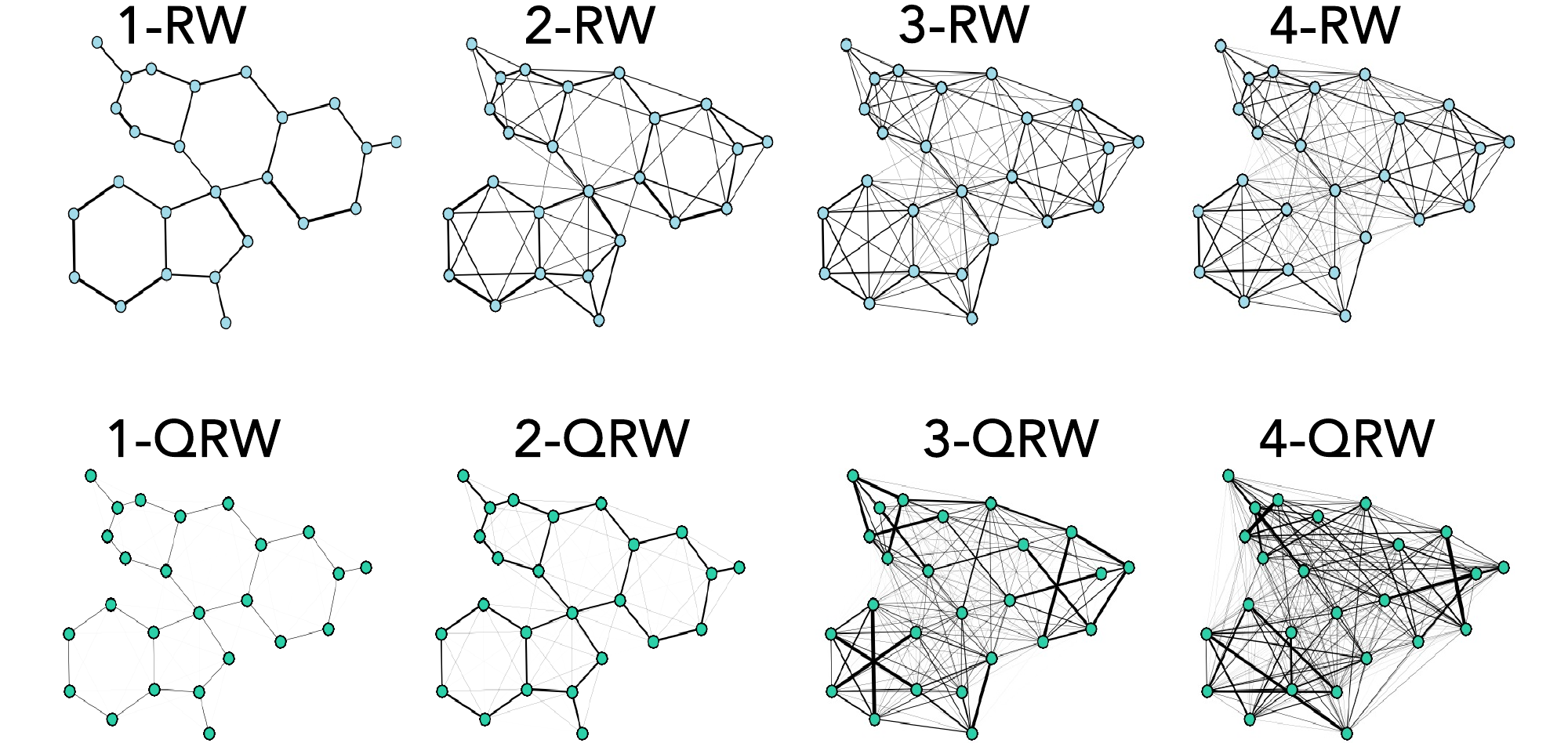}
\caption{
Visualization of RRWP (RW) and CQRW (QRW) steps for the fluorescein molecule that highlights different substructures (edge thickness). Note stronger long-range correlations (thicker edges between distant nodes) in 4th step of QRW versus RW.
}
\label{fig:rw}
\end{figure}

\section{Methods}
\label{sec:method_theory}

In this section, we outline the process of mapping graphs to a quantum state of a QPU as used in \cite{Henry21, thabet2022extending, Verdon19}. 
To extract graph features, we introduce correlators and define the concept of the ground state for a quantum graph representation. Finally, we explore an alternative approach for extracting graph features using quantum random walks (QRW) and their advantages over classical analogues. Figure~\ref{fig:schema} shows the summary of our hybrid approach in extracting positional encodings from graphs. We discuss the various steps in more detail below. 

\subsection{Quantum Graph Machine Learning}
\textbf{The graph as a quantum state.}
\label{sec:QGML}
We explain in this subsection how to create a quantum state that contains relevant information about the graph. More details about quantum information processing can be found in \citep{nielsen2002quantum}.
We associate a graph $\Graph (\vertices, \edges)$, to a quantum state $\ket{\psi_{\Graph}}$ of $|\vertices|$ qubits containing information about $\Graph$ via a hamiltonian $\Ham_\Graph$ of the form
$ \Ham_\Graph = \Sum_{(i, j) \in \edges} \Ham_{ij}$ where $\Ham_{ij}$ is an Pauli string acting non-trivially on $i$ and $j$ only.
We will be focusing on the Ising hamiltonian $\Ham^{I} = \sum_{(i, j) \in \edges} Z_i Z_j$ and  the XY hamiltonian $\Ham^{XY} = \sum_{(i, j) \in \edges} X_i X_j + Y_i Y_j$.
We will note $\ket{0}$ and $\ket{1}$ the two eigenstates (or eigenvectors) of $Z$ with respective eigenvalues 1 and -1, and we will use $\left\{\ket{\bold{b}}=\bigotimes_{i=1}^N\ket{b_i}\right\}_{\bold{b}\in \{0,1\}^N}$ as a basis of the $2^N$-dimensional space of quantum states.
We consider the quantum state obtained by alternated action of $p$ layers of $\Ham_\Graph$ and a \textit{mixing} hamiltonian  $\Ham_M$ (that doesn't commute with $\Ham_\Graph$, for instance $\Ham_M\propto\sum_i Y_i$)
\begin{gather}
\label{eqn:quantum graph state}
\ket{\psi_{\Graph}(\boldsymbol{\theta})} = \Prod_{k=1}^p\left(\e^{-i \Ham_M \theta_k}\e^{-i \Ham_\Graph t_k}\right)\e^{-i \Ham_M \theta_0}\ket{\psi_0},
\end{gather}
where $\boldsymbol{\theta} = (\theta_0, t_0, \theta_1, t_1, \dots \theta_p)$ is a real vector of parameters. 
The choice of these states is motivated by their similarity with the {\it Trotterized} dynamics of several quantum systems\citep{suzuki1976generalized}. Other states, including ones generated with analog evolution of quantum systems \cite{henriet2020quantum} can also be considered.

\textbf{Correlation.}
The correlations (or \textit{correlators}) $C_{ij}$ of local operators $\Obs_i$ and $\Obs_j$ acting respectively on qubits $i$ and $j$ can be defined either as the expectation value of their product $\langle\Obs_i\Obs_j\rangle$, or their covariance $\langle\Obs_i\Obs_j\rangle-\langle\Obs_i\rangle\langle\Obs_j\rangle$ (note that the orders matters if $\Obs_i$ and $\Obs_j$ don't commute).
In the rest of the paper, we will indifferently call correlation the two former expressions, and give precisions when necessary.
We will be focusing on the case where $\Obs_i$ is a Pauli string of length 1 (i.e., $X_i$, $Y_i$ or $Z_i$).

\textbf{Ground state.}
\label{sub:ground_state}
The ground state (GS) of a system is defined as the lowest-energy eigenstate of its hamiltonian (when it is degenerate, one considers the \textit{ground state manifold} $\mathbb{H}_{GS}$).
GSs are in the general case hard to compute classically \cite{schollwock2008quantum}.
With that in mind, we call {\it ground state of the graph} the state 
$
    |\psi_{GS}\rangle = \frac{1}{\sqrt{|\mathbb{H}_{GS}|}}\sum_{\bold{b} \in \mathbb{H}_{GS}}|\bold{b}\rangle
$. 
More details can be found in the appendix \ref{subsec:ground_state}.

\textbf{Classical and Quantum Walks.}
Quantum walks, as introduced by~\citep{aharonov1993quantum}, differ fundamentally from classical random walks by evolving through unitary processes, allowing for interference between different trajectories. 
This difference, related to the evolution of (real-valued classical) probabilities and (complex-valued quantum) amplitudes, leads to significant differences between dynamics of classical and quantum walks that can provide novel graph features which can be used as positional encodings (see Fig.~\ref{fig:rw}).
Quantum walks manifest in two primary types: \textit{continuous-time quantum walks} (CQRW) ~\citep{farhi1998quantum, rossi2017continuous} and \textit{discrete-time quantum walks} (DQRW) \citep{lovett2010universal}. 
In CQRW, the evolution of amplitudes $\langle i | \psi_\Graph({t})\rangle$ is given by $i \frac{\mathrm{d}}{\mathrm{d} t}\langle i | \psi_\Graph({t})\rangle=\sum_j\langle i|\Ham_\Graph| j\rangle\langle j | \psi_\Graph(t)\rangle$ where $| i \rangle$ is a computational basis state.
The connection to a classical CRW can be made essentially by replacing the quantum amplitudes with classical probabilities, which we detail in the first section of the appendix.
For an overview, refer to~\citep{kempe2003quantum}.
In Sec.~\ref{subsec:kqrv},  we define the particular cases of quantum walks studied in this work. More details can be found in the appendix \ref{subsec:qwalks}.

\subsection{Positional encodings with quantum features}
In this section, we detail our proposals to incorporate quantum features in GNN models and discuss the potential benefits and drawbacks. We focus on two types of encoding: the first uses the ground state of the graph as defined in section ~\ref{sub:ground_state}, the second uses the XY hamiltonian on the $k$-particles subspace (detailed in \ref{subsec:kqrv}). Both methods use a quantum state that is hard to prepare in the general case, therefore we expect to get features that are not available with classical algorithms. 


\subsubsection{Eigenvectors of the correlation on the ground state.}
\label{subsec:corr}
We propose to use the correlation matrix $C_{ij} = \langle Z_iZ_j\rangle$ on the ground state of the graph defined in Sec.~\ref{sub:ground_state}.
Since this matrix is symmetric with nonnegative eigenvalues, it can formally be used in the same place as the Laplacian matrix in graph learning models.
Hence, we use the eigenvectors of this correlation matrix in the same way Laplacian eigenvectors (LE) are used in other architectures of graph transformers.
Instead of taking the eigenvectors with the lowest eigenvalues as for the Laplacian eigenmaps, we take the ones with highest eigenvalues, since they are the ones in which most of the information about the correlation matrix is contained.
We expect to face the same challenges due to the sign ambiguity \citep{dwivedi2021graph, kreuzer2021rethinking}, and to implement the same techniques to alleviate them \citep{lim2022sign}.\\

\subsubsection{$k$-particles quantum random walks ($k$-QRW).}
\label{subsec:kqrv}

In this subsection, we introduce the $k$-particles (or walkers) random walk positional encoding that can be obtained using $\Ham^{XY}$. 
We denote by$\mathbb{H}_k$ the $k$-particles subspace (\textsl{i.e.} the Hilbert space obtained as the span of states $\ket{\bold{b}}$ of Hamming weight $k$, noted $\ket{i_1\ldots i_k}$, parameterized by $k$ integers $i_1\ldots i_k\in\{0, 1\}^k$).
It is a well-known property that $\Ham^{XY}$ stabilizes each of the $\mathbb{H}_k$s and we denote by $\Ham^{XY}_{k}$ the XY hamiltonian restricted to $\mathbb{H}_k$~\cite{Henry21}.  
$\Ham^{XY}_{k}$ can be seen as the adjacency matrix of a graph called the $k$ \textit{occupation graph}. 
Therefore, a quantum evolution of $\Ham^{XY}_{k}$ can be seen as a quantum walk on the $k$ occupation graph. 
We use the hamiltonian $\Ham^{XY}_{k}$ to prepare a quantum state as in equation \ref{eqn:quantum graph state} that will represent a superposition over all $k$-tuples of nodes, and we measure observables for each pair of nodes which will give edge features. 
A complete simulation of the XY hamiltonian evolution is impractical for graphs of more than 20 nodes, therefore in this work we restrict ourselves to the simulation of $\Ham^{XY}_{1}$ and $\Ham^{XY}_{2}$. 
In the following, we give details about the features that were implemented in the experiments.

\textbf{Continuous quantum random walk.}
For a 1-particle QRW, we calculate the probability $[X^{(1)}(t)]_{ij} = |\bra{j}e^{-i\Ham^{XY}_1t}\ket{i}|^2$ to find particle at node $j$ coming from node $i$ after time $t$.
Similarly for a 2-particle QRW, we calculate $[X^{(2)}(t)]_{ij} = |\bra{ij}e^{-i\Ham^{XY}_2t}\ket{\psi_i}|^2$, where $\ket{i,j}\in\Ham_{2}^{XY}$ is the state with walkers at nodes $i$ and $j$ and $\ket{\psi_i}\in\Ham_{2}^{XY}$ the initial state.
As choices for the initial distribution, we propose to use some localised state $\ket{\psi_\text{init}}\propto \ket{i j}$, or the uniform distribution over all pairs of nodes $\ket{\psi_\text{init}}\propto\sum_{(i,j)\in\vertices^2|i\neq j}\ket{i j}$, or the uniform distribution over the edges of the original graph  $\ket{\psi_\text{init}}\propto\sum_{(i,j)\in\edges}\ket{ij}$.
From these we obtain the positional encodings using $\mathbf{P}_{ij}=[I,X^{n_w}(t_1),X^{n_w}(t_2)...X^{n_w}(t_K)]_{ij}$, where $n_w=1,2$ is the number of walkers.

\textbf{Quantum-inspired random walks.}
We propose a discrete version of the quantum features described above, where we consider powers of the hamiltonian $(\Ham^{XY})^p$ for integers $p$ instead of continuous evolutions as explained in the previous paragraph. 
The discrete powers are not implemented natively on a quantum computer, hence the name \textit{quantum-inspired}. They are however directly comparable to the RRWP scheme of \cite{ma2023graph}, and they are cheaper to compute than the continuous quantum random walks.
We consider a \textit{discrete} 2-particle quantum-inspired RW (2-QiRW) encoding that reads
$\mathbf{P}_{ij} = \left[\bra{ij}((D_2^{XY})^{-1}\Ham_{2}^{XY})^k\ket{\psi_\text{init}} | k\in[0,K]\right]_{ij}$ where $D_2^{XY}$ is the diagonal matrix sum of the rows of $\Ham_{2}^{XY}$.

\section{Theory}
\label{theory_section}
We present below our main theoretical results. 
The proofs for all the original statements made in this section can be found in appendix \ref{apdx:theory}. 
We mean by $X\sqsubset Y$ that algorithm $X$ is weaker than algorithm $Y$ or that $Y$ can obtain a more refined node partition than $X$ in the context of node coloring. 
$X\not\sqsubset Y$ means that there exists graphs for which $X$ is stronger than $Y$ and vice-versa in distinguishing ability. 

\subsection{QWs and their relationship with the WL test}
In this section, we present some elements that help us to lay the groundwork for understanding the theoretical expressiveness of the approach developed in this paper. 
Since the seminal work of \citep{morris2019weisfeiler}, it has become standard practice to bound this theoretical expressiveness by the various variants of the WL test. 
We propose the same here to better understand the advantages and limitations of quantum computation in such context. 
First, we position our current results on Hamiltonian evolution by the XY model in the landscape of the WL test and its variants (propositions \ref{pr:1QW-uni}, proposition \ref{pr:kQW} \&  theorem \ref{th:1QW-loc}). We then focus on strongly regular graphs (SRGs) (propositions \ref{pr:GD-RRWP_SRG} and \ref{pr:GD-Ising}), which, as we describe next, constitute a relevant data set for the analysis of expressiveness.
\begin{proposition}
\label{pr:1QW-uni}
When the input state of the unitary is a uniform superposition of delocalized states, the following holds : $1$-QW $\sqsubset$ 1-WL.
\end{proposition} 
This result is mainly due to the fact that we start from a delocalized state (uniform superposition), and we can actually show that it is more efficient to start from a localized state: 
\begin{proposition}
\label{th:1QW-loc}
 When the input state of the unitary is a uniform superposition or localized states the following holds: 1. $1$-QW $\sqsubset$ 2-WL \:
     \& 2. $1$-QW $\not\sqsubset$ 	 1-WL
\end{proposition}
Both results presented so far are limited to a single quantum walker, but it's possible -- and arguably no more expensive on quantum computers-- to consider walks involving a larger number of walkers. For that, we simply need to prepare the initial state in a different configuration that allows such an analogy \cite{Henry21}. In such case, it is possible to link it to the $\delta$-$k$-$LWL$ test, a variant of the WL test presented in \cite{Morris2020}:
\begin{proposition}
    \label{pr:kQW} 
 When the input state of the unitary is a superposition state, the following holds :
      $k$-QW $\sqsubset$ $k$-$\delta$-LWL.
\end{proposition}
However, this result does not allow us to make a conclusive comparison between $k$-QW and $k$-WL. 
In a related paper \cite{Morris2020}, the authors also introduce an augmented version of $k$-$\delta$-LWL,denoted $k$-$\delta$-LWL$^+$, of which the $k$-$\delta$-LWL is a particular case. They then show that both $k$-$\delta$-WL and $k$-WL are less powerful than $k$-$\delta$-LWL$^+$. The definition of a hierarchical rule to precisely position the $k$-QW in comparison with the $k$-WL is not provided here and is part of our future work. We do, however, have a series of experimental results for a further comparison, in the context of a particular type of graphs, the strongly regular graphs, that we define next. 
\subsection{QWs on SRGs}
\label{sect:theory_srg}
\begin{definition}
A strongly regular graph (SRG), noted $srg(\nu , k, \lambda, \mu)$, is a graph with $\nu$ vertices of fixed degree $k$, such that every pair of adjacent vertices have a fixed number $\lambda$ of common neighbors, and every pair of non-adjacent vertices have a fixed number $\mu$ of common neighbors. 
\label{def:SRGs}
\end{definition} 
Each tuple $(\nu , k, \lambda, \mu)$ defines a family of SRGs, and it is possible to find multiple non isomorphic graphs within the same family \cite{srgs}. 
We chose to work on these particular graphs because their regularity makes them especially difficult to distinguish within the same family. For instance, \cite{zhang2023rethinking} provide a worst-case analysis for the GD-WL test, a provably more powerful version of $1$-WL 
, in the case of distance-regular graphs. Their examples include Rook’s and Shrikhande graphs, both of diameter 2. We know on the other hand that a SRG is a distance-regular graph with a diameter $2$ when $\mu \neq 0$~\cite{biggs1993algebraic}. 
In \cite{ma2023graph}, proposition 3.2., the authors show that GD-WL with RRWP distance is strictly more powerful than GD-WL with shortest path distance. They test their approach on 2 distance regular graphs. In the following, we propose to analyze the distinguishability of SRGs through the GD-WL test with RRWP distance, as well as a test involving a particular case of Hamiltonian evolution by the Ising model, constituting one of the rare configurations where it is possible to extract generic formulas without any costly simulation or quantum computing. 
\begin{proposition}
    \label{pr:3WL_SRG} 
\cite{pmlr-v139-bodnar21} It requires a $3$-WL test or higher to distinguish two non-isomorphic strongly regular graphs from the same family. 
\end{proposition} 
This result highlights the difficulty of the task, as a $3$-WL test requires overlapping information from all the triplets in the graph, and is therefore costly in terms of memory and time as the size of the graph analyzed increases. We continue in the same line with the following result: 
\begin{proposition}
    \label{pr:GD-RRWP_SRG} 
GD-WL with RRWP distance cannot, even with eigen-decomposition of the distance matrix, distinguish non isomorphic SRGs from the same family. 
\end{proposition} 
This extends to the RRWP distance the results recovered in \cite{zhang2023rethinking}, in which they show the same for shortest path distance and resistance distance.
We provide in the next section a set of experiments as empirical evidence that show that in some cases,  a GD-WL with correlations  on $2$-QW distinguishes SRGs. 
\subsection{Empirical study : Ising and XY models for the distinguishability of SRGs}
\begin{figure}
\captionsetup[subfigure]{labelformat=empty}
    \centering
    \includegraphics[width=\linewidth]{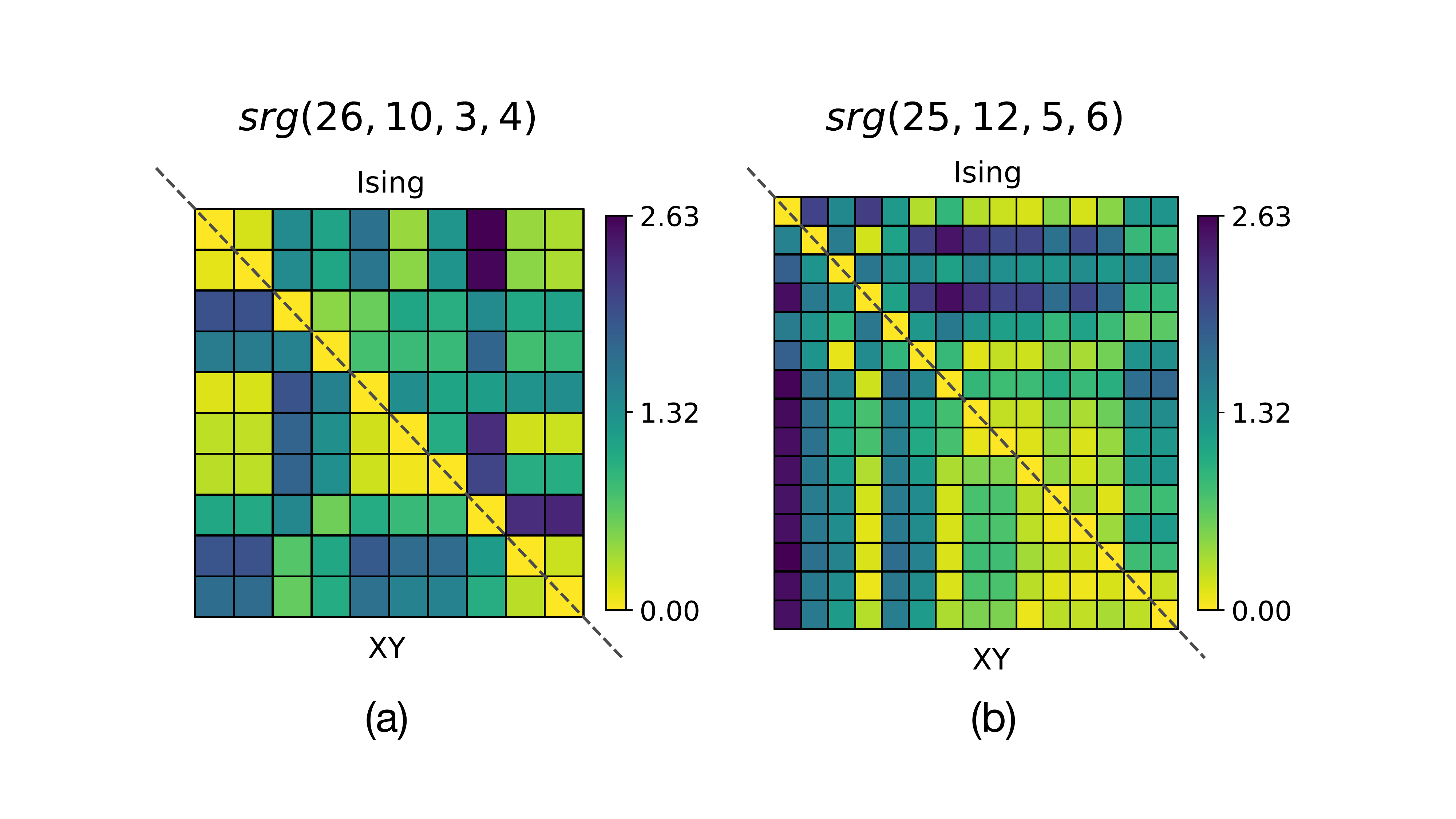}
    \caption{Normalized values of Ising correlations (upper triangular part) and XY model correlations (lower triangular part) for (a) the family $srg(26,10,3,4)$ that includes 10 non-isomorphic graphs, and (b) the family of $srg(25,12,5,6)$ that includes 15 non-isomorphic graphs.}%
    \label{fig:distinguish_SRGs}
\end{figure}

So far we only focused on the XY hamiltonian, as it offers a nice basis for theoretical analysis. This is not the case for another widely used model in many-body physics : the Ising model \textit{cf.} section \ref{sec:QGML}. As described in \cite{Henry21}, the XY hamiltonian preserves the space of the states with the same number of occupied states, which allows the analogy between classical and quantum walks. Such an invariant is not available for the Ising model, and the counterpart for augmenting the number of walkers for higher expressiveness in the case of the Ising model comes from raising the value of $p$ in equation \ref{eqn:quantum graph state}, which can be seen as piling a larger number of "layers" in the hamiltonian evolution described there \cite{Verdon19}. In this framework, we limit ourselves to the case of 1 and 2 bodies observables for many practical reasons. First they allow us to recover one and two dimensional tensors to describe graphs, which is convenient for learning tasks. They also allow us to construct the covariance matrix of such observables. This matrix can then be seen as the Gram matrix of a quantum graph kernel, which makes it suitable for comparison with the different variants of the GD-WL test. Finally, it is possible to derive a formal expression for one and two body observables resulting from an evolution for $p =1 $ in the general case, but not for larger values of $p$, which requires costly simulations or a quantum computer. 
\begin{proposition}
    \label{pr:GD-Ising} 
For any quantum state representing a strongly regular graph following an evolution from a local, uniform-field Ising Hamiltonian, the total occupation observable, as well as the linear (one body) and quadratic (two bodies) local occupation observables do not allow to distinguish two non isomorphic SRGs from the same family. 
\end{proposition} 
This result shows that the case in which it is possible to derive formal expressions actually fails to distinguish non isomorphic SRGs. Figure \ref{fig:distinguish_SRGs} shows the results of the distinguishability in a set of 2 families, with 25 and 26 nodes. We compute both Ising and XY-hamiltonian evolutions, for $ p = 2$ in the former and for a $2$-QW in the latter. 
We use the following permutation invariant measure to compute the distance between a pair of graphs  $
    d(G_i,G_j) = \frac{1}{2}\Sum_{k = 1}^{\nu^2}||S(C(G_i)) - S(C(G_j)||_k
$ where $C$ is the correlation matrix, $S(M)$ a function that receives a matrix $M \in \mathbb{R}^{N \times N}$ as input and returns a vector $m \in \mathbb{R}^{N^2}$ containing the sorted elements of $M$. Using this formula, a non- zero value implies that the two graphs are not isomorphic, but the opposite is not necessarily true. We can see from this figure that it is possible to distinguish these graphs. We also ran experiments to verify that the distance between any of these graphs and a set of 5 randomly selected isomorphic counterparts is zero as expected.This shows empirically that for certain data sets, a two layers Ising evolution and a $2$-QW are strictly more powerful than $2$-WL. We also run the GD-WL test with the edge features $S(M)$ and we obtain that all pairs of graphs are successfully distinguished.
Finally, it's important to point out that the classical $k$-WL test requires comparisons between pairs of subsets of nodes of size $k$, rendering its complexity to at least $N^k$, $N$ being the size of the graph. On the other hand, for quantum evolutions, the complexity of the algorithm is characterized by the number of shots that need to be measured in order to reconstruct the distribution of the desired observable. The number of shots 
is $\mathcal{O}(\frac{1}{\epsilon^2})$ for a precision up to $\epsilon$ \cite{huang2020predicting}. This number does not increase with the number of layers in the Ising hamiltonian case, where we only have a linear increase in the evolution time (which is short in practice), or with the number of quantum walkers that only depends on the initial input state preparation. This gives our approach an attractive potential for quantum advantage, albeit limited to datasets on which large values of $k$ in the $k$-WL test are relevant. This also assumes that the $k$-QW as well as the $k$-layers Ising, are both strictly more powerful than the $k$-WL for any values of $k$. This property has been demonstrated in some cases for $k=1$ and observed for $k=2$ in the case of SRGs, but not yet demonstrated in the general case and for any value of $k$. 

\section{Experiments}
\label{sec:expe}

\subsection{Experiments on RW models}
In this subsection, we test concatenating the QRW encodings to the RRWP in the GRIT model \citep{ma2023graph}. We compute the (continuous) 1-CQRW for $K$ random times and the discrete 2-QiRW for $K$ steps.  
Those encodings are computed numerically since they are still tractable for graphs below 200 nodes compared to the higher order $k$-QiRW ones. 
We benchmark our method on 7 datasets from \citep{dwivedi2020benchmarking}, following the experimental setup of \citep{rampavsek2022recipe} and \citep{ma2023graph}. 
Our method is compared to many other architectures and the results directly taken from \citep{ma2023graph}. 
We do not perform an extensive hyperparameter search for each architecture and only run ourselves the GRIT model by taking the same hyperparameters as the authors. 
The experiments are done by building on the codebase of \citep{ma2023graph} which is itself built on \citep{rampavsek2022recipe}. 
More details about the protocol can be found in \ref{app:qrw_experiment}, and more details about the datasets can be found in Appendix \ref{app:datasets}.
The results are included in Table \ref{tab:exp_2rw}. 
Our methods performs better on ZINC, MNIST and CIFAR10 than all others, and comes second for PATTERN and CLUSTER. We also benchmark our methods on large-scale datasets, ZINC-full (a bigger version of ZINC \citep{irwin2012zinc}) and PCQM4MV2  \citep{hu2021ogb}. For these datasets, we run a variety of models (GINE, GatedGCN, and GRIT) with different position encodings (LE, RRWP, 2-QiRW, and a mix of RRWP and 2-QiRW. The results are reported figure \ref{fig:large_datasets} and the full numbers are table \ref{tab:large_datasets_numbers} in the appendix.
Quantum features perform better for all models in the case of ZINC-full and for some models of PCQM4Mv2. All hyperparameters for this sections are reported in the appendix in tables \ref{tab:hyperparams_grit} and \ref{tab:hyperparams_large}. 

\begin{table*}[h]
    \centering
    \caption{
    Test performance in five benchmarks from \citep{dwivedi2020benchmarking}. 
    We show the mean $\pm$ s.d. of 4 runs with different random seeds as in \citep{ma2023graph}. Highlighted are the top \first{first}, \second{second}, and \third{third} results. 
    Models are restricted to $\sim500K$ parameters for ZINC, PATTERN, CLUSTER  $\sim 100K$ for MNIST and CIFAR10. We compare our model to our run of GRIT and indicate the results obtained by the authors for information. 
    Figures other than the last 3 lines are taken from \citep{ma2023graph}. Models in {\bf bold} are our models.
    }
    {\scriptsize
    \begin{tabular}{lccccc}
    \toprule
       \textbf{Model}  &  \textbf{ZINC} & \textbf{MNIST} & \textbf{CIFAR10} & \textbf{PATTERN} & \textbf{ CLUSTER} \\
       \cmidrule{2-6} 
       & \textbf{MAE}$\downarrow$  &  \textbf{Accuracy}$\uparrow$ & \textbf{Accuracy}$\uparrow$ & \textbf{Accuracy}$\uparrow$ & \textbf{Accuracy}$\uparrow$ \\
       \midrule
GIN  & $0.526 \pm 0.051$ & $96.485 \pm 0.252$ & $55.255 \pm 1.527$ & $85.387 \pm 0.136$ & $64.716 \pm 1.553$ \\
GatedGCN & $0.282 \pm 0.015$ & $97.340 \pm 0.143$ & $67.312 \pm 0.311$ & $85.568 \pm 0.088$ & $73.840 \pm 0.326$ \\
EGT & $0.108 \pm 0.009$ & \second{$\mathbf{98.173 \pm 0.087}$} & $68.702 \pm 0.409$ & $86.821 \pm 0.020$ &  \third{$\mathbf{79.232 \pm 0.348}$} \\
 GPS & {${0.070} \pm {0.004}$} & $98.051 \pm 0.126$ & {${72.298 \pm 0.356}$} & $86.685 \pm 0.059$ & {${78.016 \pm 0.180}$} \\
 GRIT (our run) & $\third{\mathbf{0.060 \pm 0.002}}$ & $\third{\mathbf{98.164 \pm 0.054}}$ & \third{$\mathbf{76.198 \pm 0.744}$} & \first{$\mathbf{90.405 \pm 0.232}$} & \first{$\mathbf{79.856 \pm 0.156}$}\\
\midrule
{\bf GRIT 1-CQRW } & \first{$\mathbf{0.058 \pm 0.002} $}& $98.108 \pm 0.111$ & \second{$\mathbf{76.347 \pm 0.704}$}  &  \third{$\mathbf{87.205 \pm 0.040}$}  & $78.895 \pm 0.1145$ \\
{\bf GRIT 2-QiRW } & \second{$\mathbf{0.059 \pm 0.004}$} &\first{$\mathbf{98.204\pm 0.048}$} & $\first{\mathbf{76.442 \pm 1.07}}$ & \second{$\mathbf{90.165 \pm 0.446}$} & \second{$\mathbf{79.777 \pm 0.171}$}\\
       \bottomrule
    \end{tabular}
    }
    \\
    \label{tab:exp_2rw}
\end{table*}

\begin{figure}[h]
     
         \includegraphics[width=0.9\columnwidth]{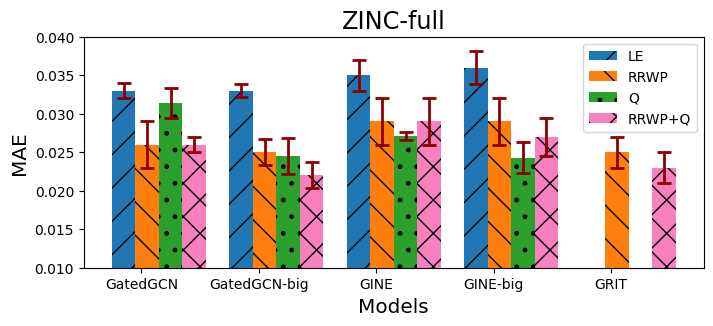}

         \includegraphics[width=0.9\columnwidth]{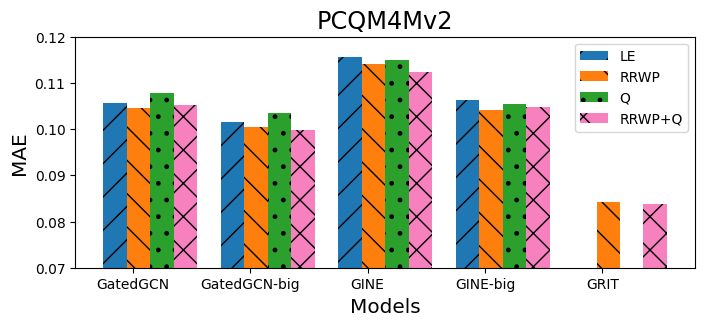}
         
        \caption{Test performance (mean absolute error) of different models with different positional encodings on large scale datasets. Q: 2-QiQRW features, RRWP+Q: RRWP and 2-QiQRW concatenated. Top : ZINC-full~. Bottom:  PCQM4Mv2. For ZINC-full, we show the mean and s.d of 4 runs with different random seeds. For PCQM4Mv2 we show the output of a single run.GRIT has 500k parameters for ZINC-full and 11.8M for PCQM4MV2. Normal models have about 200k parameters and big models about 700M.}
\label{fig:large_datasets}
\end{figure}

\subsection{Synthetic experiments}

\begin{figure}[h]
     
         \includegraphics[width=0.9\columnwidth]{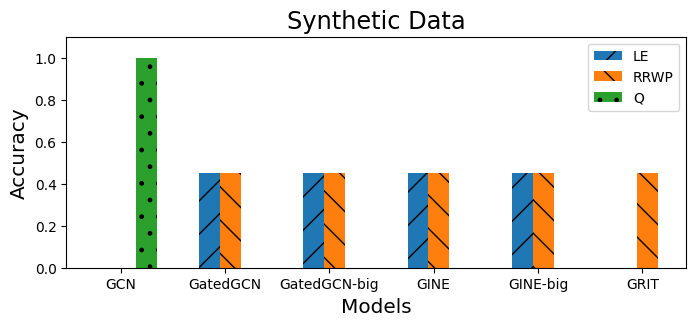}
         
        \caption{Results on synthetic data. We show the accuracy on the test set. Q: eigenvectors of the correlation on the ground state. GCN model has about 4k parameters, GRIT model has 500k parameters, other normal models have about 10k parameters, big models have about 60k parameters.}
\label{fig:synth_data}
\end{figure}

In this section, we provide one example of dataset with a binary graph classification task for which the use of the correlation matrix on the ground state as defined in \ref{subsec:corr} is more powerful than other commonly used features like the laplacian eigenvectors or RRWP. 
The idea is to construct graphs that will exhibit very different Ising ground states but similar spectral properties or random walk transition probabilities. We illustrate the differences between the encodings in Appendix \ref{app:syth_experiment}. We train classical models like GINE, GatedGCN, and GRIT on these datasets with LEs and RRWP as node features and edge features, and we compare it to a simple GCN model with the eigenvectors of the correlation matrix as node features. 
More details on the protocols can be found in \ref{app:syth_experiment} We also benchmark the GRIT model with RRWP. The results are shown in table \ref{fig:synth_data}. The quantum encoding models achieve 100\% accuracy whereas all other classical models achieve 45\% accuracy.

\subsection{Discussion}

We performed several experiments comparing the quantum encodings to the classical ones. 
Including the quantum walk features into state-of-the-art models improves their performances on most of the datasets tested. 
It is not surprising that the method works well for datasets such as ZINC for which random walks are known to provide relevant features \citep{rampavsek2022recipe}. 
We limited ourselves to versions of quantum features that are efficiently computable and we were able to show a small gain in performance compared to state-of-the-art models. 
It is then plausible that using quantum features that cannot be classically accessible could lead to a greater improvement of models, if quantum hardware can be made widely available. 
We were able to engineer an artificial dataset for which classical approaches fail to perform the associated binary classification tasks and our quantum encoding perfectly realizes it, even with only 4k parameters where classical models have between 10k and 500k.
\section{Conclusion}
In this paper, we have investigated how quantum computing architectures can be used to construct new families of positional encodings for graph neural networks.
This study involved measuring observables like correlations and probabilties for a quantum system with a hamiltonian that has the same topology as the graph of interest.
We then integrated these observables as positional encodings and used them in different classical graph neural network architectures.
We proved that some positional encodings that use quantum features are theoretically more expressive than ones based on simple random walk and laplacian eigenvectors, on strongly regular graphs.
Our experiments show that state-of-the-art models can already be enhanced with restricted quantum features that are classically efficient to compute.
This study provides strong indications that the full leverage of quantum hardware can lead to the development of high-performance architectures for certain tasks.
Quantum simulation and analog quantum computing platforms are particularly suited to the type of time-dependent Hamiltonians we described here. In particular, Neutral Atom quantum hardware \citep{henriet2020quantum, aquila} implement these natively (including the XY model \citep{XY_rydberg}) and they are becoming usable through cloud access.
Furthermore, we can create artificial classification tasks that are easily solvable with quantum-enhanced models while classical models fail.
While it remains unclear whether an actual performance improvement will be observed on real-world use cases, and whether this improvement will justify the cost of using quantum hardware in the future., 
the results we obtain show that quantum-enhanced GNNs are a promising family of models that could be fully exploited with near-term quantum hardware.

\section{Impact statement}
This paper presents work whose goal is to advance the field of Machine Learning. There are many potential societal consequences of our work, none which we feel must be specifically highlighted here.

\section{Acknowledgements}
We thank Romain Fouilland and Shenyang Huang for useful discussions. We thank Shaheen Acheche for useful advice for writing the manuscript.


\bibliography{refs}
\bibliographystyle{icml2024}

\newpage
\appendix
\onecolumn

\section{Methods}

In this section, we present the details about classical and quantum random walks.

\subsection{Differences between classical and quantum random walks}
\label{subsec:qwalks}
In this section, we make the connection between (continuous) classical and quantum walks explicit.
In CRWs, the probability of a walker being at vertex $i$ and time $t$ is represented as $p_i(t)$, which follows the differential equation $\frac{\mathrm{d}}{\mathrm{d}t}p_i(t) = - \sum_{j} G_{ij} p_j(t)$. Here, the infinitesimal generator $G_{ij} = -\gamma$ if an edge exists between nodes $i$ and $j$, and $0$ otherwise, with diagonal elements $G_{ii} = k_i \gamma$ determined by the node degree $k_i$. 
Considering now a quantum evolution with a graph Hamiltonian $\Ham_\Graph$, given a $2^N$-dimensional Hilbert space of $N$ qubits, 
the Schrödinger equation which governs the evolution of a quantum state $|\psi_{\Graph}\rangle$ when projected onto a state $|i\rangle$ is given as
\begin{align}
\underbrace{
i \frac{\mathrm{d}}{\mathrm{d} t}\langle i | \psi_\Graph({t})\rangle=\sum_j\langle i|\Ham_\Graph| j\rangle\langle j | \psi_\Graph(t)\rangle}_{\rm{Quantum}} \longleftrightarrow 
\underbrace{\frac{\mathrm{d}}{\mathrm{d} t}p_i(t) = - \sum_{j} G_{ij} p_j(t).}_{\rm{Classical}}
\end{align}
Note the similarity between the differential equations of CQRW and CRW. 
A quantum analogue of CRW can be obtained by taking $\langle i|\Ham_\Graph| j\rangle = G_{ij}$. 
 The probabilities are preserved as the sum of amplitude squared, $\sum_i |\langle i | \psi_{\Graph}(t)\rangle|^2 = 1$, in the quantum case, instead of $\sum_i p_i(t)=1$ in the classical case. 
Using this formalism, any quantum evolution can be thought of as a CQRW~\citep{childs2002example}. 
Notably, quantum walks have demonstrated exponential hitting time advantage for graphs like hypercubes~\citep{kempe2002quantum} and glued binary trees~\citep{Childs_2003}.
These results have been recently extended for more general hierarchical graphs~\citep{balasubramanian2023exponential}.
For an overview, refer to~\citep{kempe2003quantum}.

\subsection{Ground state.}
\label{subsec:ground_state}
The ground state of a system is defined as the lowest-energy eigenstate of its hamiltonian (when it is degenerate, one considers the \textit{ground state manifold} $\mathbb{H}_{GS}$).
Ground state properties are widely studied in many-body physics and their properties depend on the topology of the graph.
Preparing this state is the purpose of quantum annealing \citep{das2008colloquium}.
When using neutral atom quantum processors \citep{henriet2020quantum}, one can natively address hamiltonians of the form $\Ham_\Graph = \sum_{(i, j) \in \edges} J_{ij}(Z_i -\alpha_i I)(Z_j-\alpha_j I)$, with $\alpha_i$ real coefficients. Its eigenstates are the basis states $\ket{\bold{b}}$ described above.
In the case where $\alpha_i=1-\delta/(2z_i)$ with $z_i=\sum_{j|(i,j)\in\edges}J_{ij}$ and $J_{ij}=1/4$, the eigenenergies (or eigenvalues) are $E(\bold{b}) = \sum_{i, j \in \edges}b_ib_j - \delta \sum_{i=1}^N b_i$. When $0<\delta<1$, this is the cost function associated with the maximum independent set problem, a NP-hard problem \citep{bookNP}. 
In the absence of degeneracy-lifting or symmetry-breaking effects, a quantum annealing scheme would prepare a symmetric, equal-weight superposition of all maximum independent sets. With that in mind, we will call {\it ground state of the graph} the state 
$
    |\psi_{GS}\rangle = \frac{1}{\sqrt{|\mathbb{H}_{GS}|}}\sum_{\bold{b} \in \mathbb{H}_{GS}}|\bold{b}\rangle
$.

\section{Theory}
\label{apdx:theory}
In this section, we present the proofs relating the expressiveness of QW with well-known WL tests and other graph invariants such as the Furer Invariant \cite{Furer2010}.

\subsection{Proof of proposition \ref{pr:1QW-uni}}
 $1$-QW $\sqsubset$ $1$-WL when the input state of the unitary is a uniform superposition state.
 \newline We show that beginning with an initial uniform superposition state, the action of ${\mathcal{H_{XY}}}$ evolution is similar to WL coloring iterations. We consider a graph $\mathcal{G}$, with vertex set $\mathcal{V}$ and consisting of $N$ nodes. Suppose we begin with 1 particle QW. The action of ${\mathcal{H_{XY}}}$ restricted to 1-particle subspace is then equivalent to action of the adjacency matrix $A$. 
Therefore we have :
\begin{equation}
\begin{split}
    e^{\ii\mathcal{H_{XY}}t}& = \Sum_{k = 0} ^{\infty} \frac{(\ii.t)^k}{k!} (H_{XY})^k \\
    & \equiv 1 + \ii At + \ii^2\frac{1}{2}A^2t^2+..
\end{split}
\end{equation}

The action of $A$ on the identity vector is simply the vector of degrees of the nodes, which is the first WL iteration. Further operating with $A$ leads to a vector where we add degrees of the neighbours of a node. Further iterations simply lead to vectors where two entries are equal if they have same degrees and equal number of neighbours also of same degrees, which corresponds to stable WL coloring. This scheme corresponds exactly to a $1$-WL test with the sum operation as the update function.
To see this, suppose we define $|I\rangle$ as the normalized vector of uniform superposition. Each entry of the vector is $\frac{1}{\sqrt{N}}$. After the action of adjacency matrix $A$ we obtain the vector $|I^{(1)}\rangle$, which will be:
\begin{equation}
	|I^{(1)}\rangle = \frac{1}{\sqrt{N}}\begin{bmatrix}
				d_1\\
				d_2\\
				..\\
				..\\
				d_N
			\end{bmatrix}
\end{equation}  
where $d_1,d_2,...,d_N$ are the degrees of the nodes. This can be interpreted as the first step of WL coloring where each node is assigned the color depending on the degree. Again applying $A$ to vector $|I^{(1)}\rangle$ we obtain the vector $|I^{(2)}\rangle$, which will be:
\begin{equation}
	|I^{(1)}\rangle = \frac{1}{\sqrt{N}}\begin{bmatrix}
				\sum_{i\in \mathcal{N}(1)}d_i\\
				\sum_{i\in \mathcal{N}(2)}d_i\\
				..\\
				..\\
				\sum_{i\in \mathcal{N}((N)}d_i
			\end{bmatrix}
\end{equation}  
where $\mathcal{N}(i)$ is the set of neighbours of node $i$. Thus in this iteration, the degrees of the neighbours of a node are aggregated \footnote{Again, we are here in a case where the aggregation function is the sum operation, which is not an injective function. This is the main reason for which the $1-QW$ is less powerful than the $1-WL$, which allows any sort of aggregation functions and most notably injective functions.}. Further steps simply lead to aggregation of neighbour properties for a given node until stable coloring is reached i.e same colored neighbours will have same entry in the vector.
It is known that $i^{th}$ entry of $A^{k}|\boldsymbol{1}\rangle$ will correspond to the number of walks of length $k$ beginning from node $i$. Every $k^{th}$ iteration induces a refinement on the previous step until convergence. It has been shown that stable WL coloring is a refinement of the partition obtained at convergence for some $k$ for $A^{k}|\boldsymbol{1}\rangle$ and is in fact equal for a large number of graphs \cite{Powers1982}.  Since the evolution due to $\mathcal{H_{XY}}$ is simply the weighted summation of these vectors, it follows that that the output vector for 1-QW with input superposition $\sqsubset$ 1-WL.

\subsection{Proof of proposition \ref{th:1QW-loc}}

$1$-QW $\not\sqsubseteq$ $1$-WL $\sqsubseteq$ $2$-WL  when the input state of the unitary is a localized state.
\newline The complete proof relies on this two following statements :
\begin{enumerate}
    \item 1-QW $\not\sqsubseteq$ 1-WL when the input state of the unitary is a general input state. \newline 
    We can see that in fact for a localized input state a single-particle, QW can be more expressive than a 1-WL test for some graphs. For example, beginning with a particle localized at one of the nodes, the action of the ${\mathcal{H_{XY}}}$ as seen above has contributions from different powers of $A$. Beginning with a localized input state, the $k^{th}$ power of $A$ tell about the number of paths of length $k$ between a set of nodes. If 2 regular graphs differ in these quantities, the set of probabilities $p_{ij}=p(i\rightarrow j)$ or probability of a particle to go from node $i$ to $j$ will be different. Thus, a single particle QW can distinguish 2 such regular graphs. However, we know that a 1-WL cannot distinguish regular graphs. Hence 1-QW $\not\sqsubseteq$ 1-WL for a general input. 

    \item 1-QW $\sqsubset$ F.I when the input state of the unitary is a localized state. \newline
    For a general input and output state, the time averaged transition probability for evolution with $\mathcal{H_{XY}}$ to go from $|\psi\rangle$ to $|\phi\rangle$ is given by:

\begin{equation}
p(|\phi\rangle \rightarrow |\psi\rangle) = lim_{T  \rightarrow \infty}\int_{0}^{T} dt |\langle \phi |e^{i\mathcal{H_{XY}}t}|\psi\rangle|^2 dt
\end{equation}
This can be further written as

\begin{equation}\label{qw_probs}
   p_{|\psi\rangle \rightarrow |\phi\rangle}(\infty) = \sum_{\lambda \in Sp({\mathcal{H_{XY}}})} |\langle \phi |P^\lambda |\psi\rangle |^2 
\end{equation}
where ${\mathcal{H_{XY}}}= \sum_{\lambda \in Sp({\mathcal{H_{XY}}})} \lambda P^{\lambda}$, $Sp({\mathcal{H_{XY}}})$ is the set of eigenvalues of $\mathcal{H_{XY}}$. For a single particle QW, this set is exactly the set of eigenvalues of $A$. $P^\lambda$ are the projectors into the eigenspace spanned by the corresponding eigenvectors. \cite{Furer2010} defines the spectral invariant based on projectors $P^{\lambda}$. We refer to this as the Furer Invariant (F.I.). 
The Furer invariant is a spectral invariant defined in terms of projectors $P^\lambda$ defined above. The projector $P^\lambda$ by definition can be written as $|e_\lambda\rangle \langle e_\lambda|$, where $|e_{\lambda}\rangle$ is the eigenvector corresponding to the eigenvalue $\lambda$. In matrix form this can be written as 
\begin{equation}
	P^\lambda = \begin{bmatrix}
			p^{\lambda}_{11}&p^{\lambda}_{11}&..&p^{\lambda}_{1n}\\
			p^{\lambda}_{21}&p^{\lambda}_{22}&..&p^{\lambda}_{2n}\\
			.&.&.\\
			.&.&.\\
			p^{\lambda}_{n1}&p^{\lambda}_{n2}&..&p^{\lambda}_{nn}\\
   \end{bmatrix}
\end{equation}
The quantities $\langle b_i|P^\lambda|b_j\rangle$ and $\langle b_i|P^\lambda|b_i\rangle$ represent the non-diagonal and the diagonal entries of the matrix $P^{\lambda}$ respectively. The $|b_i\rangle$ represent the coordinate basis vectors where a vector $|0_1,0_2..,1_i,0..,0\rangle $ would represent the node $i$. The quantity $\langle b_i|P^\lambda|b_j\rangle$ represents the product of angles made by the coordinate basis vectors corresponding to node $i,j$ with the eigenspace defined by $\lambda$. Similarly $\langle b_i|P^\lambda|b_i\rangle$ represents the angle made by the coordinate basis vector of node $i$ with the eigenspace defined by $\lambda$. These quantities along with the spectra of the adjacency matrix define the Furer invariant. Expanding each of the vectors $|\phi\rangle$ and $|\psi\rangle$ in terms of the eigenvectors, it can be seen that the probability $p(i \rightarrow j)$ can be computed using the Furer invariant. Thus, the set of probabilities is weaker than the Furer invariant.

Thus, we see that 1-QW$\sqsubset$F.I. In \cite{Rattan2023}, the authors prove that F.I$\sqsubset$2-WL. Therefore 1-QW$\sqsubset$ 2-WL. These inequalities along with Proposition 1.1 tell us that 1-QW $\not\sqsubseteq$ 1-WL and 1-QW $\sqsubseteq$ 2-WL.
    
\end{enumerate}
\subsection{Proof of proposition \ref{pr:kQW}}

$k$-QW $\sqsubset$ to  $k$-$\delta$-LWL for uniform input superposition state.
\newline 
For a general $k$-particle case, the WL coloring for a $k$-particle occupation graph is the same as $k$-$\delta$-LWL coloring as defined in \cite{Morris2020}. This is because the colors of the $k$-tuples are updated only based on the local neighbourhood, as is the case for the $k$-$\delta$-LWL test. The k-particle occupancy graph $G^{k}$, where every node now labelled by a $k$-tuple represents the presence of $k$ particles at that node, is constructed in the following way.: We begin with a $k$-tuple , $(v_1,v_2,..v_{i},v_{i+1},..v_k)$. An adjacent $k$-tuple is found by  replacing $v_i$ with a node $w_i$ such that $w_{i}\in\mathcal{N}(v_i)$. In a $k$-$\delta$-LWL test, colors of a particular $k$-tuple are updated based on colors of the $\it{local}$ neighborhood, in which two $k$-tuples $(v_1,v_2,..v_{i},v_{i+1},..v_k)$ and $(v_1,v_2,..w_{i},v_{i+1},..v_k)$ are local neighbours only if $w_{i}\in\mathcal{N}(v_i)$ in the original graph. Thus the updates for both the cases are equivalent. Analogously to the $1$-QW case, we see therefore that $k$-QW $\sqsubset$ to  $k$-$\delta$-LWL.

\subsection{Proof of proposition \ref{pr:GD-RRWP_SRG}}
One can show \citep{shiau2003physically, gamble2010two} that for strongly regular graphs, the powers of the adjacency matrix $A$ can be expressed as $$A^n = \alpha_n I + \beta_n J + \gamma_n A$$ where $\alpha_n$, $\beta_n$, $\gamma_n$ only depend on $N, k, \lambda, \mu$. $I$ is the identity matrix, $J$ is the matrix full of 1s.\newline The degree matrix $D$ is also equal to $kI$, then $(D^{-1}A)^n = A^n/k^n$. Hence the information about distance contained in $\mathbf{P}_{uv}$ for strongly regular graphs is the same as in their adjacency matrices. Therefore, for strongly regular graphs, the GD-WL with RRWP test is equivalent to the WL test.
The F.I. consists of eigenvalues and eigenvectors. It has been shown that F.I is stronger than $1$-WL, but weaker than $2$-WL \cite{Rattan2023}. Therefore, $GD$-WL along with eigenvectors cannot distinguish SRGs, which require $3$-WL or higher.

\subsection{proof of proposition \ref{pr:GD-Ising}}
In this proof, we provide the full calculation for 

\begin{itemize}
    \item The average value of the global occupation observable $\langle \hat n (t) \rangle_\Graph = \langle \Sum_{v \in \mathcal{V}} \hat n_v(t)\rangle_\Graph $
    \item the correlation matrix for the local occupation observables $C_{uv}(t) = \langle \hat n_{u}(t) . \hat n_{v}(t) \rangle_\Graph -  \langle \hat n_{u}(t)\rangle_\Graph  \langle \hat n_{v}(t) \rangle_\Graph$
\end{itemize}
We provide for each of these cases a formulation involving the corresponding graph quantities. In each of these steps, the goal is to show that it is not a powerful tool to distinguish graphs, and more specifically, that it fails to distinguish non isomorphic strongly regular graphs from the same family. 

We start by defining the following quantities : 

\begin{equation}\hat n_{v} = \frac{1}{2}(\mathbb{I} - \hat Z_v)
\end{equation} 
and
\begin{equation}
 \hat Z_v = \displaystyle\bigotimes_{u<v}\mathbb{I}\otimes\hat Z\otimes\displaystyle\bigotimes_{u>v}\mathbb{I}
 \end{equation} 
We also define 
\begin{equation}
    \langle \hat{\mathcal{O}} \rangle_\Graph = \langle \psi_\Graph | \hat{\mathcal{O}} | \psi_\Graph \rangle
    \label{eq:val_moy}
 \end{equation} 
where 
\begin{equation}
   |\psi_\Graph \rangle = \mathcal{U}_\Graph |\psi_0 \rangle  = \mathcal{U}_\Graph \bigotimes_{v =1}^{n}|0 \rangle 
 \end{equation} 
For our case, the evolution operator $ \mathcal{U}_\Graph $ can be written as : 
\begin{equation}
\mathcal{U}_\Graph(\vartheta, t) = \e^{i \Ham_M \vartheta}\e^{-i \Ham_I t} \e^{-i \Ham_M \vartheta}
\end{equation}
where, for a graph $\Graph = (\mathcal{V},\edges)$ we have  
\begin{equation}
    \Ham_{I} = \Sum_{(v,u)\in\edges} \hat n_v \hat n_u + \Sum_{v \in \vertices} \hat n_v 
\end{equation}
		 
 and \footnote{Physically, this describes the same phenomenon as if we had chosen a mixing hamiltonian with operators $X_i$, up to a gauge change. Which leads to similar results that do not affect the statement in the corresponding proposition.}
 \begin{equation}
     \Ham_M=  \Sum_{v\in\mathcal{V}} \hat Y_v
 \end{equation}
 
This corresponds to a single layer Ising evolution, preceded and followed by an evolution through the mixing hamiltonian $\Ham_M$ with opposite signs. The motivation for switching signs in the mixing hamiltonian parameter $\vartheta$ is to allow us to write  $\mathcal{U}_\Graph(\vartheta, t) = \mathcal{U}^{\dag}_M(\vartheta)\cdot\mathcal{U}_I(t) \cdot \mathcal{U}_M(\vartheta)$ which is equal to the identity when $t = 0$. 
\newline Now let us compute the quantities $ \langle \hat n_{v} \hat n_{u}  \rangle_\Graph$ and $\langle \hat n_{v} \rangle_\Graph$

We start by defining the following pulse operator, that will be useful later in this calculation :
\begin{equation}
    \hat{P}^\dagger_\vartheta = \exp\left(-\ii\,\vartheta \Sum_v \hat Y_v\right)=\Prod_v\underbrace{\left[\Cos\,\mathbb{I}+\Sin\left(\hat\sigma_v^+-\sigma_v^-\right) \right]}_{\hat{P}^\dagger_{\vartheta,v}}
\end{equation}		

where we introduce the spin operators $\hat\sigma^+$ and $\hat\sigma^-$, such that
			
			\begin{equation}
				\begin{array}{cc}
					\hat\sigma^+|0\rangle = |1\rangle,	&	\hat\sigma^+|1\rangle = 0,\\
					\hat\sigma^-|0\rangle = 0,			&	\hat\sigma^-|1\rangle = |0\rangle,
				\end{array}
			\end{equation}
   such that
		       $\hat X = (\hat\sigma^++\hat\sigma^-)/2,\hat Y =-\ii(\hat\sigma^+-\hat\sigma^-)/2$ and
			$\hat Z =\hat\sigma^+\hat\sigma^-$.
\subsubsection{State preparation}
    \begin{enumerate}
        \item We start with the system in 
            \begin{equation}
                |{\bf 0}\rangle = \displaystyle \bigotimes_v |0\rangle.
            \end{equation}
        \item We then apply a pulse of angle $\vartheta$ to get the system in the state
            \begin{equation}
                |\psi_{\vartheta}(0)\rangle = \hat{P}^\dagger_\vartheta|{\bf 0}\rangle=(\Cos)^N\Sum_{\sigma\in\{0,1\}^N} (\Tan)^{n_\sigma}|\sigma\rangle.
            \end{equation}
        \item We then let the system evolve with the Ising Hamiltonian, so that
            \begin{equation}
                |\psi_{\vartheta}(t)\rangle = \e^{-\ii \Ham_I t}|\psi_{\vartheta}(0)\rangle=(\Cos)^N\Sum_{\sigma\in\{0,1\}^N} (\Tan)^{n_\sigma}\e^{-\ii E^{(I)}_\sigma}|\sigma\rangle.
            \end{equation}
        \item Finally, the inverse pulse is then applied, to get
            \begin{equation}
                |\psi_{f,\vartheta}(t)\rangle = \hat{P}_\vartheta|\psi_{\vartheta}(t)\rangle =(\Cos)^N\Sum_{\sigma\in\{0,1\}^N} (\Tan)^{n_\sigma}\e^{-\ii E^{(I)}_\sigma t}\left(\Prod_{v=1}^N\hat{P}_{\vartheta,v}\right)|\sigma\rangle.
            \label{eq:ref_eq}
            \end{equation}
    \end{enumerate}

\subsubsection{Computation of the total occupation}
We first start by providing a full expression, in terms of graph quantities, for a (graph level) observable that we call the global occupation : 
\begin{equation}
    \hat{n}= \Sum_v\hat{n}_v =\Sum_v \hat\sigma^+_v\hat\sigma^-_v,
\end{equation}

Such that, when transposed in the expression provided in equations \ref{eq:ref_eq} and \ref{eq:val_moy} we have : 

\begin{equation}
     \langle \hat n(t) \rangle = (\Cos)^{2N}\Sum_{\sigma,\sigma'\in\{0,1\}^N} \e^{-\ii \left(E^{(I)}_\sigma-E^{(I)}_{\sigma'}\right)t}
                    \underbrace{(\Tan)^{n_\sigma+n_{\sigma'}}
                    \langle\sigma'|\left(\Prod_{v'=1}^N\hat{P}^\dagger_{\vartheta,v'}\right)\hat n \left(\Prod_{v=1}^N\hat{P}_{\vartheta,v}\right)|\sigma\rangle}
                    _{:=f_{\sigma\sigma'}(\vartheta)}.
\end{equation}

We then insert an identity $\mathbb{I} = \Sum_{\mu \in \{0,1\}^N} |\mu\rangle\langle\mu|$, so that
\begin{equation}
    \label{eq:noft_lambda}
    \langle \hat n(t) \rangle= (\Cos)^{2N}\Sum_{\sigma,\sigma'\in\{0,1\}^N} \e^{-\ii \left(E^{(I)}_\sigma-E^{(I)}_{\sigma'}\right)t}(\Tan)^{n_\sigma+n_{\sigma'}}\Sum_{\mu \in \{0,1\}^N} 
                    n_\mu 
                    \underbrace{\langle\sigma'|\left(\Prod_{v'=1}^N\hat{P}^\dagger_{\vartheta,v'}\right)|\mu\rangle
                    \langle\mu|\left(\Prod_{v=1}^N\hat{P}_{\vartheta,v}\right)|\sigma\rangle}_{=
                    \Prod_{v=1}^N
                    \langle\sigma'_v|\hat{P}^\dagger_{\vartheta,v}|\mu_v\rangle
                    \langle\mu_v|\hat{P}_{\vartheta,v}|\sigma_v\rangle}.
\end{equation}

Let us now compute an expression for the matrix elements appearing in \eqref{eq:noft_lambda} :
\begin{equation}
    \begin{array}{lcl}
    \lambda_{\mu_v\sigma_v}(\vartheta) &:=& \langle\mu_v|\hat{P}_{\vartheta,v}|\sigma_v\rangle \\
                                &=& \Cos\, \langle\mu_v|\sigma_v\rangle +  \Sin\, \langle\mu_v|\hat\sigma^-_v|\sigma_v\rangle -  \Sin\, \langle\mu_v|\hat\sigma_v^+|\sigma_v\rangle\\
                                &=& \Cos\, \delta_{\mu_v,\sigma_v} +  \Sin\, \left(\delta_{\mu_v,\sigma_v-1} -  \delta_{\mu_v,\sigma_v+1}\right)
    \end{array}
\end{equation}

%
The following table gives the possible values for the product $\lambda^*_{\mu_v\sigma'_v}(\vartheta)\lambda_{\mu_v\sigma_v}(\vartheta)$ :
        \begin{equation}
            \label{eq:table}
            \begin{array}{|c|c|}\hline
                (\mu_v,\sigma_v,\sigma'_v)	&	\lambda^*_{\mu_v\sigma'_v}(\vartheta)\lambda_{\mu_v\sigma_v}(\vartheta)\\\hline\hline
                (0,0,0),(1,1,1)				&	\Cos^2\\
                (0,1,1),(1,0,0)				&	\Sin^2\\
                (0,1,0),(0,0,1)				&	\Cos\Sin\\
                (1,1,0),(1,0,1)				&	-\Cos\Sin\\\hline
            \end{array}
        \end{equation}

In order to further explicit \eqref{eq:noft_lambda}, we split the set of all vertices into $\mathcal{O}_{\sigma\sigma'} = \{ v | \sigma_v = \sigma'_v=1\}$, $\mathcal{Z}_{\sigma\sigma'} = \{ v | \sigma_v = \sigma'_v=0\}$ and 
$\Delta_{\sigma\sigma'} = \{ v | \sigma_v \neq \sigma'_v\}$, respectively containing $n_0, n_1$, and $n_{\neq}$ vertices, so that 
\begin{equation}
        \Prod_{v=1}^N\lambda^*_{\mu_v\sigma'_v}(\vartheta)\lambda_{\mu_v\sigma_v}(\vartheta) =
    \Prod_{v\in\mathcal{O}_{\sigma\sigma'}}\underbrace{\lambda^*_{\mu_v\sigma'_v}(\vartheta)\lambda_{\mu_v\sigma_v}(\vartheta)}_{=(\Cos)^{2{\mu_v}}(\Sin)^{2-2{\mu_v}}}
    \Prod_{v\in\mathcal{Z}_{\sigma\sigma'}}\underbrace{\lambda^*_{\mu_v\sigma'_v}(\vartheta)\lambda_{\mu_v\sigma_v}(\vartheta)}_{=(\Cos)^{2-2{\mu_v}}(\Sin)^{2{\mu_v}}}
    \Prod_{v\in\Delta_{\sigma\sigma'}}\underbrace{\lambda^*_{\mu_v\sigma'_v}(\vartheta)\lambda_{\mu_v\sigma_v}(\vartheta)}_{\Cos\Sin(-1)^{{\mu_v}}}.
\end{equation}	

We then split $\mu$ in $\mu = \mu^{(1)} \cup \mu^{(0)} \cup \mu^{(\neq)}$, where $\mu^{(1)}=\{\mu_v|v\in\mathcal{O}_{\sigma\sigma'}\}$, $\mu^{(0)}=\{\mu_v|v\in\mathcal{Z}_{\sigma\sigma'}\}$
and $\mu^{(\neq)}=\{\mu_v|v\in\Delta_{\sigma\sigma'}\}$. We can then write
\begin{equation*}
        f^\mu_{\sigma\sigma'}(\vartheta)=\Prod_{v=1}^N\lambda^*_{\mu_v\sigma'_v}(\vartheta)\lambda_{\mu_v\sigma_v}(\vartheta) =\underbrace{
    (\Sin)^{2n_1}(\Tan)^{-2n_{\mu^{(1)}}}\times
    (\Cos)^{2n_0}(\Tan)^{2n_{\mu^{(0)}}}\times
    (\Cos)^{n_{\neq}}(\Sin)^{n_{\neq}}(-1)^{n_{\mu^{(\neq)}}}}
    _{ (\Cos)^{2N}(\Tan)^{n_{\neq}+2n_1}(\Tan)^{2(n_{\mu^{(0)}}-n_{\mu^{(1)}})} (-1)^{n_{\mu^{(\neq)}}} }.
\end{equation*}	
Noting that $\Sum_{\mu\in\{0,1\}^m} A(n_{\mu}) = \Sum_{p=0}^m\binom{m}{p}A(p)$ we can write
\begin{equation*}
    \Sum_{\mu \in \{0,1\}^N} n_\mu f^\mu_{\sigma\sigma'}(\vartheta) = \underbrace{(\Cos)^{2N}(\Tan)^{n_{\neq}+2n_1}
    \underbrace{
    \Sum_{p_1=0}^{n_1}\binom{n_1}{p_1}
    \underbrace{
    \Sum_{p_0=0}^{n_0}\binom{n_0}{p_0}(\Tan)^{2(p_0-p_1)}
    \underbrace{\Sum_{p_{\neq}=0}^{n_{\neq}}\binom{n_{\neq}}{p_{\neq}}
    \left(p_1+p_0+p_{\neq}\right)
     (-1)^{p_{\neq}}}_{=(p_1+p_0)\delta_{0,n_{\neq}} - \delta_{1,n_{\neq}} }}
     _{=\left[1+(\Tan)^{2}\right]^{n_0}\left[\delta_{0,n_{\neq}}\left(p_1+n_0\frac{(\Tan)^{2}}{1+(\Tan)^{2}}\right)  - \delta_{1,n_{\neq}}\right]      }
     }
     _{=\left[1+(\Tan)^{2}\right]^{n_0}\left[1+(\Tan)^{-2}\right]^{n_1}\left\{\delta_{0,n_{\neq}}\left[n_0(\Sin)^2+n_1\frac{(\Tan)^{-2}}{1+(\Tan)^{-2}}\right]-\delta_{1,n_{\neq}}\right\}}}_
     {
     (\Cos)^{2N}(\Tan)^{n_{\neq}}\left[1+(\Tan)^{2}\right]^{n_0+n_1}\left\{\delta_{0,n_{\neq}}\left[n_0(\Sin)^2+n_1(\Cos)^2\right]-\delta_{1,n_{\neq}}\right\}
     }
\end{equation*}

Since $n_0+n_1+n_{\neq}=N$, 
\begin{equation}
    \label{eq:intermediate1}
     (\Cos)^{2N}(\Tan)^{n_{\neq}}\left[1+(\Tan)^{2}\right]^{n_0+n_1} = \frac{(\Tan)^{n_{\neq}}}{\left[1+(\Tan)^{2}\right]^{n_{\neq}} }=(\Cos\Sin)^{n_{\neq}},
\end{equation}

so that
    
\begin{equation}
    \Sum_{\mu \in \{0,1\}^N} n_\mu f^\mu_{\sigma\sigma'}(\vartheta) =
        \delta_{0,n_{\neq}}\left[n_0(\Sin)^2+n_1(\Cos)^2\right]-\delta_{1,n_{\neq}}\Cos\Sin
\end{equation}

The sum in \eqref{eq:noft_lambda} then only runs on pairs of configuration $(\sigma,\sigma')$ that differ by at most 1 element. Let us simplify those terms.

\begin{itemize}
    \item {\bf $n_{\neq} = 0$} \\
        Then $\sigma=\sigma'$, so that $n_\sigma=n_{\sigma'}=n_1=N-n_0$. 
        \begin{equation}
            f_{\sigma\sigma}(\vartheta) =(\Tan)^{2n_\sigma}\left[(N-n_\sigma)(\Sin)^2+n_\sigma(\Cos)^2\right]
        \end{equation}
    \item {\bf $n_{\neq} = 1$} 
        \begin{itemize}
            \item{$n_{\sigma'} = n_\sigma+1 $ } \\
                Then $n_1=n_\sigma=N-n_0-1$, so that
            \begin{equation}
                \label{eq:sigleqsigp}
                f_{\sigma\sigma'}(\vartheta) = -(\Tan)^{2n_\sigma+1}\Cos\Sin = -(\Tan)^{2n_\sigma}\sin^2\vartheta
            \end{equation}									
            \item{$n_{\sigma'} = n_\sigma-1 $ } \\
                Then $n_1=n_{\sigma'}=N-n_0-1$, so that
            \begin{equation}
                \label{eq:siggeqsigp}
                f_{\sigma\sigma'}(\vartheta) = -(\Tan)^{2n_\sigma-1}\Cos\Sin = -(\Tan)^{2n_\sigma}\cos^2\vartheta
            \end{equation}
        \end{itemize}
        
        {\bf Remark :} Exchanging $\sigma$ and $\sigma'$ in either of the previous equations \eqref{eq:sigleqsigp} or \eqref{eq:siggeqsigp} yields the other one,
        ensuring that $\langle \hat n(t) \rangle$ is real. For example, if $n_{\sigma'} = n_\sigma+1 $
        \begin{equation}
            f_{\sigma'\sigma}= -(\Tan)^{2n_\sigma'}\cos^2\vartheta = -(\Tan)^{2n_\sigma+2}\cos^2\vartheta =  -(\Tan)^{2n_\sigma}\sin^2\vartheta = f_{\sigma\sigma'}
        \end{equation}
\end{itemize}

						For a given configuration $\sigma$, we note $\Delta_{\sigma,v}=E_\sigma-E_{\sigma'(v)}$ where $\sigma'(v)$ is the configuration obtain from $\sigma$ by flipping
						$\sigma_v$. The total occupation is then
						
						\begin{equation}
							\langle \hat n(t) \rangle = \Sum_{\sigma\in\{0,1\}^N}\left\{
							f_{\sigma\sigma}(\vartheta)+
							\Sum_{v\in\mathcal{Z}_\sigma}\e^{-\ii\Delta_{\sigma,v}t}f_{\sigma\sigma'(v)}(\vartheta)+
							\Sum_{v\in\mathcal{O}_\sigma}\e^{-\ii\Delta_{\sigma,v}t}f_{\sigma\sigma'(v)}(\vartheta)
							\right\}
						\end{equation}
						
						Let us put this in a form that makes it more explicitly real. We start with
						
						\begin{equation}
							\Sum_{\sigma\in\{0,1\}^N}
							\Sum_{v\in\mathcal{Z}_\sigma}\e^{-\ii\Delta_{\sigma,v}t}f_{\sigma\sigma'(v)}(\vartheta)\underbrace{=}_{\nu=\sigma'(v)}
							\Sum_{\nu\in\{0,1\}^N}	\Sum_{v\in\mathcal{O}_\nu}\e^{\ii\Delta_{\nu,v}t}f_{\nu'(v)\nu}(\vartheta)=
							\Sum_{\nu\in\{0,1\}^N}	\Sum_{v\in\mathcal{O}_\nu}\e^{\ii\Delta_{\nu,v}t}f_{\nu\nu'(v)}(\vartheta).
						\end{equation}
						
						The total occupation becomes
						
						\begin{equation}
							\langle \hat n(t) \rangle = \Sum_{\sigma\in\{0,1\}^N}\left\{
							f_{\sigma\sigma}(\vartheta)+
							2\Sum_{v\in\mathcal{O}_\sigma}\cos\left(\Delta_{\sigma,v}t\right)f_{\sigma\sigma'(v)}(\vartheta)
							\right\}.
						\end{equation}
						
						The first term can be rewritten as
						\begin{equation}
							\Sum_{\sigma\in\{0,1\}^N} 	f_{\sigma\sigma}(\vartheta) = 
								(\Cos)^{2N}\Sum_{p=0}^N\binom{N}{p} (\Tan)^{2p}\left[(N-p)\sin^2\vartheta+p\cos^2\vartheta\right]=
								2N\sin^2\vartheta\cos^2\vartheta
						\end{equation}

						The total occupation becomes
						
\begin{equation}
\label{eq:noftsig}
\langle \hat n(t) \rangle =2N\sin^2\vartheta\cos^2\vartheta-
2(\Cos)^{2N+2}\Sum_{\sigma\in\{0,1\}^N}(\Tan)^{2n_\sigma}
\Sum_{v\in\mathcal{O}_\sigma}\cos\left(\Delta_{\sigma,v}t\right)
\end{equation}

\paragraph{Remark :} for $n_\sigma = 0$, the term in the sum vanishes $\left(\Sum_{v\in\mathcal{O}_\sigma}\cos\left(\Delta_{\sigma,v}t\right)=0\right)$.

At $t=0$, this becomes
\begin{equation}
n(0) =2N\sin^2\vartheta\cos^2\vartheta-
2(\Cos)^{2N+2}\Sum_{\sigma\in\{0,1\}^N}(\Tan)^{2n_\sigma}n_\sigma
=2N\sin^2\vartheta\cos^2\vartheta-
2(\Cos)^{2}
N\frac{\tan^2\vartheta}{1+\tan^2\vartheta}=0
\end{equation}

\subsubsection{Expression in terms of graph}
	For an induced subgraph $\Graph_\sigma$ of $\Graph$ and a vertex $v\in\vertices_\sigma$, we note $\kappa_\sigma(v)$ the degree of $v$ in $\Graph_\sigma$, and $\kappa(v)$ its degree in $\Graph$. 
	The energy difference between the Ising configurations corresponding to $\Graph_\sigma$ and to $\Graph_\sigma \setminus {v}$ is  $\kappa(v)$.
	The total occupation can then be expressed as 
						
\begin{equation}
\label{eq:noftgraph}
\langle \hat n(t) \rangle_ \Graph =2N\sin^2\vartheta\cos^2\vartheta-
2(\Cos)^{2N+2}\Sum_{\Graph_\sigma\subset\Graph}(\Tan)^{2|\vertices_\sigma|}
\Sum_{v\in\vertices_\sigma}\cos\left(\kappa_{\sigma}(v)t\right).
\end{equation}

We can also express the sum as

\begin{equation}
	\Sum_{\Graph_\sigma\subset\Graph}(\Tan)^{2|\vertices_\sigma|}
\Sum_{v\in\vertices_\sigma}\cos\left(\kappa_{\sigma}(v)t\right) =
\Sum_{v\in\vertices}
\Sum_{\vertices_\sigma\ni v}(\Tan)^{2|\vertices_\sigma|}
\cos\left(\kappa_{\sigma}(v)t\right).
\end{equation}
						
We note $V(v) = \{v'\in\vertices|(v,v')\in\edges\}$ the set of all neighbours of $v$ in $\Graph$, as well as $V_\sigma(v) = \{v'\in\vertices_\sigma|(v,v')\in\edges_\sigma\}$ the set of all neighbours of $v$ in $\Graph_\sigma$, so that\footnote{$|V_\sigma(v)| = \kappa_\sigma(v)$} 
\begin{equation}
\label{eq:cos}
\Sum_{v\in\vertices}
\Sum_{\vertices_\sigma\ni v}(\Tan)^{2|\vertices_\sigma|}
\cos\left(\kappa_{\sigma}(v)t\right)=
\Sum_{v\in\vertices}\Sum_{\omega=0}^{\kappa(v)}
\Sum_{\substack{\vertices_\sigma\ni v|\\\kappa_\sigma(v)=\omega }}(\Tan)^{2|\vertices_\sigma|}
\cos\left(\omega t\right).
\end{equation}

For a given degree $\omega\leq \kappa(v)$, and a given integer $n\geq0$, there are $\binom{N-1-\kappa(v)}{n}\binom{\kappa(v)}{\omega}$ induced subgraphs $\Graph_\sigma$ of size $n+\omega+1$, containing $v$, and in which
$v$ has degree $\omega$. We can then write

\begin{equation}
\Sum_{v\in\vertices}\Sum_{\omega=0}^{\kappa(v)}
\Sum_{\substack{\vertices_\sigma\ni v\\\kappa_\sigma(v)=\omega }}(\Tan)^{2|\vertices_\sigma|}
\cos\left(\omega t\right)=
\Sum_{v\in\vertices}
\underbrace{\Sum_{\omega=0}^{\kappa(v)}\binom{\kappa(v)}{\omega}\cos\left(\omega t\right)(\Tan)^{2\omega+2}}_
	{(\Tan)^2\Re\left\{\left(1+\e^{\ii  t}(\Tan)^2\right)^{\kappa(v)}\right\}}
\underbrace{\Sum_{n=0}^{N-1-\kappa(v)}\binom{N-1-\kappa(v)}{n}(\Tan)^{2n}}_
		{\left(1+(\Tan)^2\right)^{N-1-\kappa(v)}}
\end{equation}

The sum in $\langle \hat n(t) \rangle$ then becomes

\begin{equation}
2(\Cos)^{2N+2}\Sum_{\Graph_\sigma\subset\Graph}(\Tan)^{2|\vertices_\sigma|}
\Sum_{v\in\vertices_\sigma}\cos\left(\kappa_{\sigma}(v)t\right) =
2\,\sin^2\vartheta\Sum_{v\in\vertices}\frac{\Re\left\{\left(1+\e^{\ii  t}\tan^2\vartheta\right)^{\kappa(v)}\right\}}{\left(1+\tan^2\vartheta\right)^{1+\kappa(v)}}.
\end{equation}

If we note $m_\Graph(\kappa)$ the number of vertices of degree $\kappa$ in $\Graph$, and $\kappa_{\max}(\Graph)$ the maximum degree of $\Graph$, 
we can express the occupation as

\begin{equation}
	\label{eq:noft}
\langle \hat n(t) \rangle_ \Graph =
2\,{\sin^2\vartheta\cos^2\vartheta}\Sum_{\kappa=0}^{\kappa_{\max}(\Graph)} m_\Graph(\kappa)\,{\Re\left\{1-\left(\cos^2\vartheta+\e^{\ii  t}\sin^2\vartheta\right)^{\kappa}\right\}}.
\end{equation}

\begin{figure}[htpb]
	\centering
	\includegraphics[width=0.6\textwidth]{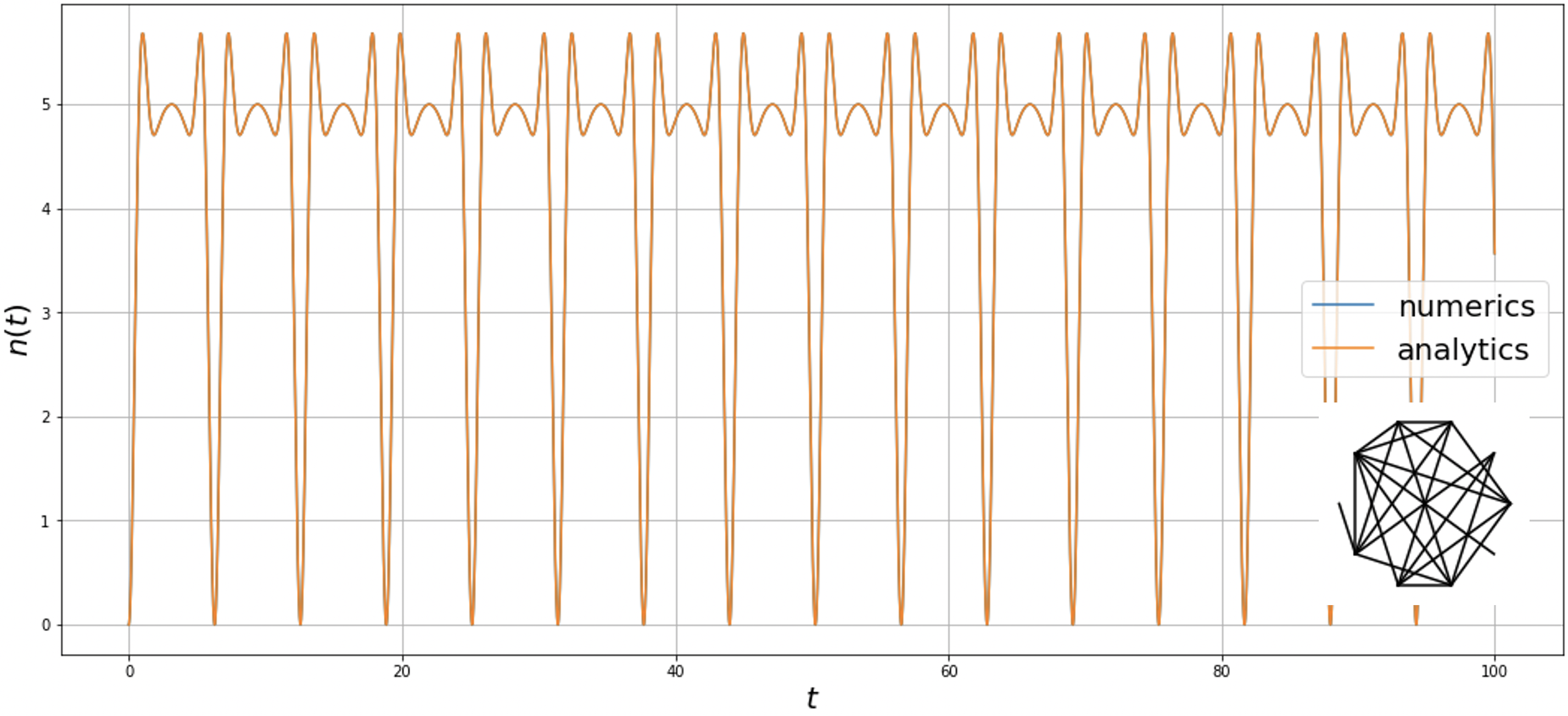}
	\caption{Comparison between \eqref{eq:noft} and numerical simulation of the dynamics, in the case of a density-density hamiltonian, on the graph shown in the inset. Both curves collapse on top
	of each other.}
	\label{fig:density}
\end{figure}

This expression is checked in Fig. \ref{fig:density}, on a random graph of 10 vertices.

\textbf{Computation for a generic Ising model}
\newline
	We now consider the case of a generic Ising model, defined by the Hamiltonian
	\begin{equation}
		\label{eq:HamIsinggen}
		\Ham_{I_{gen}} = \Sum_{(i,j)\in\edges} J_{ij} \hat n_i \hat n_j + \Sum_{v \in \vertices} h_v \hat n_v.
	\end{equation}
	In this case, \eqref{eq:noftsig} becomes
	\begin{equation}
\langle \hat n(t) \rangle =2N\sin^2\vartheta\cos^2\vartheta-
2(\Cos)^{2N+2}\Sum_{v\in [0,N-1]}\Sum_{\substack{\sigma\in\{0,1\}^N\\ \sigma_v=1}}(\Tan)^{2n_\sigma}\cos\left(\Delta_{\sigma,v}t\right).
	\end{equation}
	We note $\mathcal{N}_\Graph(v) = \left\{v'\in\vertices|(v,v')\in\edges\right\}$ the neighbourhood of $v$ in $\Graph$, and split the Ising configuration $\sigma$ into
	$\sigma = \underbrace{\left\{ \sigma_{v'} | v'\in \mathcal{N}_\Graph(v)\right\}}_{\sigma_{\mathcal{N}_\Graph(v)}}
			\cup \{\sigma_v\}
			\cup
			\underbrace{\left\{ \sigma_{v'} | v'\not\in \mathcal{N}_\Graph(v)\cup\{v\}\right\}}_{\bar\sigma_{\mathcal{N}_\Graph(v)}}$,
	so that
	\begin{equation}
		\Sum_{\substack{\sigma\in\vertices\\ \sigma_v=1}}(\Tan)^{2n_\sigma}\cos\left(\Delta_{\sigma,v}t\right) = 
		\Sum_{\sigma_{\mathcal{N}_\Graph(v)}\in\{0,1\}^{\kappa(v)}}(\Tan)^{2(1+n_{\sigma_{\mathcal{N}_\Graph(v)}})}\cos\left(\Delta_{\sigma,v}t\right)
		\underbrace{\Sum_{\sigma_{\mathcal{N}_\Graph(v)}\in\{0,1\}^{N-1-\kappa(v)}}(\Tan)^{2n_{\bar\sigma_{\mathcal{N}_\Graph(v)}}}}_
		{\left(1+\tan^2\vartheta\right)^{N-\kappa(v)-1}}.
	\end{equation}
	We can then write
	\begin{equation}
		2(\Cos)^{2N+2}\Sum_{v\in \vertices}\Sum_{\substack{\sigma\in\{0,1\}^N\\ \sigma_v=1}}(\Tan)^{2n_\sigma}\cos\left(\Delta_{\sigma,v}t\right)=
		2\cos^2\vartheta\sin^2\vartheta\Sum_{v\in \vertices}\Sum_{\sigma\in\{0,1\}^{\kappa(v)}}\frac{(\Tan)^{2n_{\sigma}}}{\left(1+\tan^2\vartheta\right)^{\kappa(v)}}\cos\left(\Delta_{\sigma,v}t\right).
	\end{equation}
	Finally, we can further simplify the second sum into
	\begin{equation}
		\Sum_{\sigma\in\{0,1\}^{\kappa(v)}}\frac{(\Tan)^{2\Sum_{i=1}^{\kappa(v)} \sigma_{i}}}{\left(1+\tan^2\vartheta\right)^{\kappa(v)}}\cos\left[(h_v+\Sum_{i=1}^{\kappa(v)} J_{v\tilde v_i}\sigma_{i})t\right] =
		\Sum_{\kappa=0}^{\kappa(v)} \frac{(\Tan)^{2\kappa}}{\left(1+\tan^2\vartheta\right)^{\kappa(v)}}\Sum_{\substack{\tilde v\in \mathcal{N}_\Graph(v)^\kappa\\\tilde v_i\neq\tilde v_j}}\cos\left[(h_v+\Sum_{i=1}^\kappa J_{v\tilde v_i})t\right].
	\end{equation}
	Or, 
	\begin{equation}
		\Sum_{\sigma\in\{0,1\}^{\kappa(v)}}\frac{(\Tan)^{2\Sum_{i=1}^{\kappa(v)} \sigma_{i}}}{\left(1+\tan^2\vartheta\right)^{\kappa(v)}}\cos\left[(h_v+\Sum_{i=1}^{\kappa(v)} J_{v\tilde v_i}\sigma_{i})t\right] =
		\Re\left\{\e^{\ii h_v t}
		\Prod_{\tilde v\in\mathcal{N}_\Graph(v)}
		\left(\frac{1+\tan^2\,\vartheta \e^{\ii J_{v\tilde v}t}}{1+\tan^2\vartheta}\right)
		\right\}.
	\end{equation}
	
	The density $\varrho_\Graph(t) = n(t)/N$ then becomes
\begin{equation}
\label{eq:noftgeneric}
\varrho_\Graph(t) =2\sin^2\vartheta\cos^2\vartheta\,\,\Sum_{v\in\vertices}\Re \left\{1-
		\e^{\ii h_v t}\Prod_{\tilde v\in\mathcal{N}_\Graph(v)}
		\left({\cos^2\vartheta+\sin^2\vartheta\;\e^{\ii J_{v\tilde v}t}}\right)
\right\}
	\end{equation}

	In this form, it is very easy to check that $n(0)=0$.

\textbf{More generic pulse}
\newline We consider here the case where the system is first subject to a $\vartheta$ pulse, then evolves freely for a duration $t$ and finally is subject to a $\varphi$ pulse. In this case the average number of excitations is
\begin{equation}
\langle \hat n(t) \rangle_ \Graph=\sin\vartheta\cos\vartheta\sin\varphi\cos\varphi\,\,\Sum_{v\in\vertices}\Re \left\{\frac{\tan\vartheta}{\tan\varphi}+
        \frac{\tan\varphi}{\tan\vartheta}+
		2\,\,\e^{\ii h_v t}\Prod_{\tilde v\in\vertices}
		\left({\cos^2\vartheta+\sin^2\vartheta\;\e^{\ii J_{v\tilde v}t}}\right)
\right\}
	\end{equation}

If we consider the case of the uniform local Ising model, \textit{i.e.} $h_v = 1 \forall v \in \mathcal{V}$ and $J_{v_1,v_2}$ is the adjacency matrix of the graph, then we can write :  

\begin{tcolorbox}[ams gather]
\label{eq:noftgeneric}
\langle \hat n(t) \rangle_ \Graph =\sin\vartheta\cos\vartheta\sin\varphi\cos\varphi\,\,\Sum_{v\in\vertices}\Re \left\{\frac{\tan\vartheta}{\tan\varphi}+
        \frac{\tan\varphi}{\tan\vartheta}+
		2\,\,\e^{\ii t} 
		\left({\cos^2\vartheta+\sin^2\vartheta\;\e^{\ii t}}\right)^{d_v}
\right\}
	\end{tcolorbox}
where $d_v$ is the degree of node $v$. We can see that this quantity is not enough to distinguish two graphs with the same degree histograms, which is the case of two non-isomorphic SRGs from the same family. this completes the first part of the proof.

\subsubsection{Computation of the local observables }
    As we have just shown, the total occupation seems to too simple of an observable to extract meaningful informations from the graph.
    To this end, we express the expectation value of a generic quadratic function of the {\sl bitstring} $\mu$
    \begin{equation}
        \varphi(\mu) = \Sum_{v\in\vertices} h_v \mu_v + \Sum_{(v,v')\in\edges} J_{vv'} \mu_v\mu_{v'}.
    \end{equation}
    
	We start from \eqref{eq:noft_lambda}, and replace $n_\mu$ by $\varphi(\mu)$
	\begin{equation}
	    \label{eq:phiaverage}
	    \begin{aligned}
	    \langle\varphi\rangle(t)=(\Cos)^{2N+3}(\Sin)^3
	                        \Sum_{\mu \in \{0,1\}^N}\varphi(\mu)
	                        \Sum_{\sigma,\sigma'\in\{0,1\}^N} &\e^{-\ii \Delta E_{\sigma\sigma'}t}(\Tan)^{n_\sigma+n_{\sigma'}}\\
							&\times\Prod_{v\in\mathcal{O}_{\sigma\sigma'}}{(\Tan)^{-2{\mu_v}}}
						\Prod_{v\in\mathcal{Z}_{\sigma\sigma'}}{(\Tan)^{2{\mu_v}}}
						\Prod_{v\in\Delta_{\sigma\sigma'}}{(-1)^{{\mu_v}}}.
	    \end{aligned}
	\end{equation}
    
    \textbf{Linear term}
    \newline 
    Let us focus on a given linear term and find an expression for $\langle \mu_v\rangle$. We start from
    \eqref{eq:phiaverage}.

    We note $\mathcal{O}_v=\left\{\sigma\in\{0,1\}^N|\sigma_v=1\right\}$ and $\mathcal{Z}_v=\left\{\sigma\in\{0,1\}^N|\sigma_v=0\right\}$ then expand $\Sum_{\sigma,\sigma'\in\{0,1\}^N}$ into

    \textit{The term for $v\in\mathcal{O}_{\sigma\sigma'}$}
    	\begin{equation}
	    \begin{aligned}
			\Sum_{\mu\in\mathcal{O}_{v}} f_{\sigma\sigma'}^\mu(\vartheta)
				&= (\Cos)^{2N}(\Tan)^{n_{\neq}+2n_1}\Sum_{p_1=0}^{\bf n_1-1} \binom{n_1-1}{p_1}(\Tan)^{-2(p_1+1)}
											\Sum_{p_0=0}^{n_0} \binom{n_0}{p_0}(\Tan)^{2p_0}
											\Sum_{p_{\neq}=0}^{n_{\neq}}(-1)^{p_{\neq}}\\
				&= (\Cos)^{2N}\delta_{\sigma\sigma'}(\Tan)^{2(n_1-1)}(1+\tan^{-2}\vartheta)^{n_1-1}(1+\tan^{2}\vartheta)^{n_0}\\
				&= (\Cos)^{2N}\delta_{\sigma\sigma'}(\Tan)^{2(n_1-1)}(\Sin)^{-2(n_1-1)}(\Cos)^{-2n_0}\\
				&= \delta_{\sigma\sigma'}\cos^2\vartheta
	    \end{aligned}
	\end{equation}
    
    \textit{The term for $v\in\mathcal{Z}_{\sigma\sigma'}$}
    	\begin{equation}
	    \begin{aligned}
			\Sum_{\mu\in\mathcal{O}_{v}} f_{\sigma\sigma'}^\mu(\vartheta)
				&= (\Cos)^{2N}(\Tan)^{n_{\neq}+2n_1}\Sum_{p_1=0}^{n_1} \binom{n_1}{p_1}(\Tan)^{-2p_1)}
				&= \delta_{\sigma\sigma'}\sin^2\vartheta
	    \end{aligned}
	\end{equation}
	
    \textit{The term for $v\in\Delta_{\sigma\sigma'}$}
    	\begin{equation}
	    \begin{aligned}
			\Sum_{\mu\in\mathcal{O}_{v}} f_{\sigma\sigma'}^\mu(\vartheta)
				&= (\Cos)^{2N}(\Tan)^{n_{\neq}+2n_1}\Sum_{p_1=0}^{n_1} \binom{n_1}{p_1}(\Tan)^{-2p_1}
											\Sum_{p_0=0}^{n_0} \binom{n_0}{p_0}(\Tan)^{2p_0}
											\Sum_{p_{\neq}=0}^{n_{\neq}-1}(-1)^{p_{\neq}+1}\\
				&= -(\Cos)^{2N}\delta_{n_{\neq}-1}(\Tan)^{1+2n_1}(\Sin)^{-2n_1}(\Cos)^{-2n_0}\\
				&= -\delta_{n_{\neq}-1}\sin\vartheta\cos\vartheta
	    \end{aligned}
	\end{equation}
	
    \textit{The sum of all terms : $  \langle n_{v}(t)\rangle_\Graph$}
    \begin{equation}
	    \begin{aligned}
    		 \langle n_{v}(t)\rangle_\Graph = (\Cos)^{2N}
			&\left[\cos^2\vartheta\Sum_{\sigma\in\mathcal{O}_{v}}\tan^{2n_\sigma}\vartheta
			+\sin^2\vartheta\Sum_{\sigma\in\mathcal{Z}_{v}}{\tan^{2n_\sigma}\vartheta}\right.\\
			&-\sin\vartheta\cos\vartheta\Sum_{\sigma\in\mathcal{O}_{v}}{(\Tan)^{2n_\sigma-1}}\e^{\ii \Delta E_{\sigma\tilde\sigma^v} t}
			-\left.\sin\vartheta\cos\vartheta\e^{\ii h_{v} t}\Sum_{\sigma\in\mathcal{Z}_{v}}{(\Tan)^{2n_\sigma+1}}\e^{-\ii \Delta E_{\sigma\tilde\sigma^v} t}
		\right]
	    \end{aligned}
    \end{equation}

    We have introduced here $\tilde\sigma^{v}$ the configuration obtained from $\sigma$ by flipping $\sigma_v$. 

We focus on the first two terms first.

    \begin{equation}
	    \begin{aligned}
			\cos^2\vartheta\Sum_{\sigma\in\mathcal{O}_{v}}\tan^{2n_\sigma}\vartheta
			+\sin^2\vartheta\Sum_{\sigma\in\mathcal{Z}_{v}}\tan^{2n_\sigma}\vartheta
			&=
			\cos^2\vartheta\Sum_{n=0}^{N-1}\binom{N-1}{n}\tan^{2(n+1)}\vartheta
			+\sin^2\vartheta\Sum_{n=0}^{N-1}\binom{N-1}{n}\tan^{2n}\vartheta\\
			&=2\sin^2\vartheta\Sum_{n=0}^{N-1}\binom{N-1}{n}\tan^{2n}\vartheta\\
			&=2\sin^2\vartheta\cos^2\vartheta(\Cos)^{-2N}.
	    \end{aligned}
    \end{equation}
    And now, the third term, using \eqref{eq:HamIsinggen}

    \begin{equation}
	    \begin{aligned}
			\cos^2\vartheta\Sum_{\sigma\in\mathcal{O}_{v}}(\Tan)^{2n_\sigma}\e^{\ii \Delta E_{\sigma\tilde\sigma^v} t}
			&=\sin^2\vartheta\,\e^{\ii h_{v} t}
			\Prod_{\tilde v\in\mathcal{N}_\Graph(v)}\left(1+\tan^2\vartheta\,\e^{\ii J_{\tilde v} t}\right)
			\Sum_{n=0}^{N-1-\kappa(v)}\binom{N-1-\kappa(v)}{n}(\Tan)^{2n}\\
			&=\sin^2\vartheta\,\e^{\ii h_{v} t}
			\Prod_{\tilde v\in\mathcal{N}_\Graph(v)}\left(1+\tan^2\vartheta\,\e^{\ii J_{v\tilde v} t}\right)
			(\Cos)^{2(1+\kappa(v)-N)}\\
			&=\sin^2\vartheta\cos^2\vartheta(\Cos)^{-2N}\,\e^{\ii h_{v} t}
			\Prod_{\tilde v\in\mathcal{N}_\Graph(v)}\left(\cos^2\vartheta+\sin^2\vartheta\,\e^{\ii J_{v\tilde v} t}\right).
	    \end{aligned}
    \end{equation}
    
    Similarly, the fourth terms is

    \begin{equation}
	    \begin{aligned}
			\sin^2\vartheta\Sum_{\sigma\in\mathcal{Z}_{v}}(\Tan)^{2n_\sigma}\e^{-\ii \Delta E_{\sigma\tilde\sigma^v} t}
			&=\sin^2\vartheta\,\e^{-\ii h_{v} t}
			\Prod_{\tilde v\in\mathcal{N}_\Graph(v)}\left(1+\tan^2\vartheta\,\e^{-\ii J_{v\tilde v} t}\right)
			\Sum_{n=0}^{N-1-\kappa(v)}\binom{N-1-\kappa(v)}{n}(\Tan)^{2n}\\
			&=\sin^2\vartheta\,\e^{-\ii h_{v} t}
			\Prod_{\tilde v\in\mathcal{N}_\Graph(v)}\left(1+\tan^2\vartheta\,\e^{-\ii J_{v\tilde v} t}\right)
			(\Cos)^{2(1+\kappa(v)-N)}\\
			&=\sin^2\vartheta\cos^2\vartheta(\Cos)^{-2N}\,\e^{-\ii h_{v} t}
			\Prod_{\tilde v\in\mathcal{N}_\Graph(v)}\left(\cos^2\vartheta+\sin^2\vartheta\,\e^{-\ii J_{v\tilde v} t}\right).
	    \end{aligned}
    \end{equation}

   Adding all those terms together, we get

\begin{equation}
 \langle n_{v}(t)\rangle_\Graph =2\sin^2\vartheta\cos^2\vartheta\,\,\Re \left\{1-
		\e^{\ii h_v t}\Prod_{\tilde v\in\mathcal{N}_\Graph(v)}
		\left({\cos^2\vartheta+\sin^2\vartheta\;\e^{\ii J_{v\tilde v}t}}\right)
\right\}
	\end{equation}

\textbf{Quadratic term}
\newline
We no focus on a given quadratic term and find an expression for $\langle \mu_{v_1}\mu_{v_2}\rangle$, where $v_1\neq v_2$. We start from again from 
    \eqref{eq:phiaverage}.

    \textit{$v_1,v_2\in\mathcal{O}_{\sigma\sigma'}$}
    
    	\begin{equation}
		\begin{aligned}
			\Sum_{\mu\in\mathcal{O}_{v_1}\cap\mathcal{O}_{v_2}} f_{\sigma\sigma'}^\mu(\vartheta)
				&= (\Cos)^{2N}(\Tan)^{n_{\neq}+2n_1}\Sum_{p_1=0}^{\bf n_1-2} \binom{n_1-2}{p_1}(\Tan)^{-2(p_1+2)}
											\Sum_{p_0=0}^{n_0} \binom{n_0}{p_0}(\Tan)^{2p_0}
											\Sum_{p_{\neq}=0}^{n_{\neq}}(-1)^{p_{\neq}}\\
				&= \cos^4\vartheta \, \delta_{\sigma\sigma'}
		\end{aligned}
    	\end{equation}
	Note that here $\forall v, \sigma_v=\sigma'_v$, and
	\begin{equation}
		(\Tan)^{n_\sigma+n_{\sigma'}} \Sum_{\mu\in\mathcal{O}_{v_1}\cap\mathcal{O}_{v_2}} f_{\sigma\sigma'}^\mu(\vartheta) =
			\cos^4\vartheta (\Tan)^{2n_\sigma} \, \delta_{\sigma\sigma'}
	\end{equation}
    
    \textit{$v_1\in\mathcal{O}_{\sigma\sigma'},v_2\in\mathcal{Z}_{\sigma\sigma'}$}
    
    	\begin{equation}
		\begin{aligned}
			\Sum_{\mu\in\mathcal{O}_{v_1}\cap\mathcal{O}_{v_2}} f_{\sigma\sigma'}^\mu(\vartheta)
				&= (\Cos)^{2N}(\Tan)^{n_{\neq}+2n_1}\Sum_{p_1=0}^{\bf n_1-1} \binom{n_1-1}{p_1}(\Tan)^{-2(p_1+1)}
											\Sum_{p_0=0}^{\bf n_0-1} \binom{n_0-1}{p_0}(\Tan)^{2(p_0+1)}
											\Sum_{p_{\neq}=0}^{n_{\neq}}(-1)^{p_{\neq}}\\
				&= \sin^2\vartheta\cos^2\vartheta  \,\delta_{\sigma\sigma'}
		\end{aligned}
    	\end{equation}
	Note that here $\forall v, \sigma_v=\sigma'_v$, and
	\begin{equation}
		(\Tan)^{n_\sigma+n_{\sigma'}} \Sum_{\mu\in\mathcal{O}_{v_1}\cap\mathcal{O}_{v_2}} f_{\sigma\sigma'}^\mu(\vartheta) =
			\cos^4\vartheta (\Tan)^{2n_\sigma+1} \, \delta_{\sigma\sigma'}
	\end{equation}

    \textit{$v_1\in\mathcal{O}_{\sigma\sigma'},v_2\in\Delta_{\sigma\sigma'}$} 
    
    	\begin{equation}
		\begin{aligned}
			\Sum_{\mu\in\mathcal{O}_{v_1}\cap\mathcal{O}_{v_2}} f_{\sigma\sigma'}^\mu(\vartheta)
				&= (\Cos)^{2N}(\Tan)^{n_{\neq}+2n_1}\Sum_{p_1=0}^{\bf n_1-1} \binom{n_1-1}{p_1}(\Tan)^{-2(p_1+1)}
											\Sum_{p_0=0}^{n_0} \binom{n_0}{p_0}(\Tan)^{2p_0}
											\Sum_{p_{\neq}=0}^{\bf n_{\neq}-1}(-1)^{p_{\neq}+1}\\
				&=  -\Sin\cos^3\vartheta  \,\delta_{n_{\neq}-1}
		\end{aligned}
    	\end{equation}
	Note that here $\forall v\neq v_2, \sigma_v=\sigma'_v$, and $\sigma_{v_2} = 1-\sigma'_{v_2}$, and
	\begin{equation}
		(\Tan)^{n_\sigma+n_{\sigma'}} \Sum_{\mu\in\mathcal{O}_{v_1}\cap\mathcal{O}_{v_2}} f_{\sigma\sigma'}^\mu(\vartheta) =
			-\cos^4\vartheta (\Tan)^{2(n_\sigma-\sigma_{v_2})}  \,\delta_{n_{\neq}-1} \, \bar\delta_{\sigma_{v_2}\sigma'_{v_2}}
	\end{equation}
    
    \textit{$v_1,v_2\in\mathcal{Z}_{\sigma\sigma'}$}
    
    	\begin{equation}
		\begin{aligned}
			\Sum_{\mu\in\mathcal{O}_{v_1}\cap\mathcal{O}_{v_2}} f_{\sigma\sigma'}^\mu(\vartheta)
				&= (\Cos)^{2N}(\Tan)^{n_{\neq}+2n_1}\Sum_{p_1=0}^{n_1} \binom{n_1}{p_1}(\Tan)^{-2p_1}
											\Sum_{p_0=0}^{\bf n_0-2} \binom{n_0-2}{p_0}(\Tan)^{2(p_0+2)}
											\Sum_{p_{\neq}=0}^{n_{\neq}}(-1)^{p_{\neq}}\\
				&= \sin^4\vartheta  \,\delta_{\sigma\sigma'}
		\end{aligned}
    	\end{equation}
	Note that here $\forall v, \sigma_v=\sigma'_v$, and
	\begin{equation}
		(\Tan)^{n_\sigma+n_{\sigma'}} \Sum_{\mu\in\mathcal{O}_{v_1}\cap\mathcal{O}_{v_2}} f_{\sigma\sigma'}^\mu(\vartheta) =
			\sin^4\vartheta (\Tan)^{2n_\sigma} \, \delta_{\sigma\sigma'} =
			\cos^4\vartheta (\Tan)^{2n_\sigma+4} \, \delta_{\sigma\sigma'}
	\end{equation}
    
    \textit{$v_1\in\mathcal{Z}_{\sigma\sigma'},v_2\in\Delta_{\sigma\sigma'}$}
    
    	\begin{equation}
		\begin{aligned}
			\Sum_{\mu\in\mathcal{O}_{v_1}\cap\mathcal{O}_{v_2}} f_{\sigma\sigma'}^\mu(\vartheta)
				&= (\Cos)^{2N}(\Tan)^{n_{\neq}+2n_1}\Sum_{p_1=0}^{n_1} \binom{n_1}{p_1}(\Tan)^{-2p_1}
											\Sum_{p_0=0}^{\bf n_0-1} \binom{n_0-1}{p_0}(\Tan)^{2(p_0+1)}
											\Sum_{p_{\neq}=0}^{\bf n_{\neq}-1}(-1)^{p_{\neq}+1}\\
				&=  -\sin^3\vartheta\Cos  \,\delta_{n_{\neq}-1}
		\end{aligned}
    	\end{equation}
	Note that here $\forall v\neq v_2, \sigma_v=\sigma'_v$, and $\sigma_{v_2} = 1-\sigma'_{v_2}$, and
	\begin{equation}
		\begin{aligned}
		(\Tan)^{n_\sigma+n_{\sigma'}} \Sum_{\mu\in\mathcal{O}_{v_1}\cap\mathcal{O}_{v_2}} f_{\sigma\sigma'}^\mu(\vartheta) &=
			-\sin^4\vartheta (\Tan)^{2(n_\sigma-\sigma_{v_2})}  \,\delta_{n_{\neq}-1} \, \bar\delta_{\sigma_{v_2}\sigma'_{v_2}}\\&=
			-\cos^4\vartheta (\Tan)^{2(n_\sigma+2-\sigma_{v_2})}  \,\delta_{n_{\neq}-1} \,\bar \delta_{\sigma_{v_2}\sigma'_{v_2}}
		\end{aligned}
	\end{equation}

    \textit{$v_1,v_2\in\Delta_{\sigma\sigma'}$}
    
    	\begin{equation}
		\begin{aligned}
			\Sum_{\mu\in\mathcal{O}_{v_1}\cap\mathcal{O}_{v_2}} f_{\sigma\sigma'}^\mu(\vartheta)
				&= (\Cos)^{2N}(\Tan)^{n_{\neq}+2n_1}\Sum_{p_1=0}^{n_1} \binom{n_1}{p_1}(\Tan)^{-2p_1}
											\Sum_{p_0=0}^{n_0} \binom{n_0}{p_0}(\Tan)^{2p_0}
											\Sum_{p_{\neq}=0}^{\bf n_{\neq}-2}(-1)^{p_{\neq}+2}\\
				&= \sin^2\vartheta\cos^2\vartheta  \,\delta_{n_{\neq}-2}
		\end{aligned}
    	\end{equation}
	Note that here $\forall v\neq v_2,v_1, \sigma_v=\sigma'_v$, and $\sigma_{v_1} = 1-\sigma'_{v_1}, \sigma_{v_2} = 1-\sigma'_{v_2}$, and
	\begin{equation}
		(\Tan)^{n_\sigma+n_{\sigma'}} \Sum_{\mu\in\mathcal{O}_{v_1}\cap\mathcal{O}_{v_2}} f_{\sigma\sigma'}^\mu(\vartheta) =
			\cos^4\vartheta (\Tan)^{2(n_\sigma-\sigma_{v_2}-\sigma_{v_2})}  \,\delta_{n_{\neq}-2} \, \bar\delta_{\sigma_{v_2}\sigma'_{v_2}} \,\bar \delta_{\sigma_{v_1}\sigma'_{v_1}}
	\end{equation}
	
    \textit{$ \langle n_{v_1}n_{v_2}(t) \rangle_\Graph$}
   \newline In the following, we will note $\tilde \sigma^i$ the configuration obtained from $\sigma$ by flipping $\sigma_{v_i}$, and 
    configuration obtained from $\sigma$ by flipping both $\sigma_{v_1}$ and $\sigma_{v_2}$. We then split the sum $\Sum_{\sigma\in\{0,1\}^N}$ into
    four terms
    \begin{equation}
    \Sum_{\sigma\in \mathcal{O}_{v_1}\cap\mathcal{O}_{v_2} } 
    + \Sum_{\sigma\in \mathcal{O}_{v_1}\cap\mathcal{Z}_{v_2} } 
    + \Sum_{\sigma\in \mathcal{Z}_{v_1}\cap\mathcal{O}_{v_2} }
    +\Sum_{\sigma\in \mathcal{Z}_{v_1}\cap\mathcal{Z}_{v_2} }
    := \mathcal{S}_{11}+\mathcal{S}_{10}+\mathcal{S}_{01}+\mathcal{S}_{00},
    \end{equation}
     and focus on each of them separately.

    \begin{equation}
    	\begin{aligned}
    	\mathcal{S}_{11}&=\Sum_{\sigma\in \mathcal{O}_{v_1}\cap\mathcal{O}_{v_2}}\Sum_{\sigma'\in\{0,1\}^N}
	\e^{-\ii \Delta E_{\sigma,\sigma'} t} (\Tan)^{n_\sigma+n_{\sigma'}} \Sum_{\mu\in\mathcal{O}_{v_1}\cap\mathcal{O}_{v_2}} f_{\sigma\sigma'}^\mu(\vartheta)  \\
				&= \cos^4\vartheta\Sum_{n=0}^{N-2} \binom{N-2}{n}(\Tan)^{2(n+2)}\\
				&- \cos^4\vartheta\Sum_{\sigma\in \mathcal{O}_{v_1}\cap\mathcal{O}_{v_2}}(\Tan)^{2(n_\sigma-1)}
						\left(\e^{-\ii \Delta E_{\sigma,\tilde\sigma^1} t}+\e^{-\ii \Delta E_{\sigma,\tilde\sigma^2} t}\right)\\
    	\end{aligned}
    \end{equation}

    We introduce $\mathcal{N}_{12}=\mathcal{N}_\Graph(v_1)\cap\mathcal{N}_\Graph(v_2)$, and 
    $\mathcal{N}_{1}=\mathcal{N}_\Graph(v_1)\backslash\mathcal{N}_{12}\backslash v_2$ (and $1\leftrightarrow2$).
    Furthermore, we also introduce 
    \begin{equation}
    	w_\vartheta( \phi, t ) = \left(\cos^2\vartheta + \sin^2\vartheta \, \e^{-\ii \phi t}\right)
    \end{equation}
    
      If $(v_1,v_2)\in\edges$, then
    \begin{equation}
    	\begin{aligned}
    	\mathcal{S}_{11}&=(\Cos)^{-2N}\sin^2\vartheta\cos^6\vartheta\left[\tan^2\vartheta
					-\e^{-\ii h_{v_1} t}\Prod_{\tilde v \in \mathcal{N}\backslash v_2}w_\vartheta( J_{v_1\tilde v}, t )
					-\e^{-\ii h_{v_2} t}\Prod_{\tilde v \in \mathcal{N}\backslash v_1}w_\vartheta( J_{v_2\tilde v}, t )
					\right.\\
					&
					\left.
					+\tan^2\vartheta \,\e^{-\ii (h_{v_1}+h_{v_2} + J_{v_1v_2}) t}
					\Prod_{\tilde v \in \mathcal{N}_1}w_\vartheta( J_{v_1\tilde v}, t )
					\Prod_{\tilde v \in \mathcal{N}_2}w_\vartheta( J_{v_2\tilde v}, t )
					\Prod_{\tilde v \in \mathcal{N}_{12}}w_\vartheta( J_{v_1\tilde v} + J_{v_2\tilde v}, t )
				\right]
    	\end{aligned}
    \end{equation}

	The generic expression is
    \begin{equation}
    	\begin{aligned}
    	\mathcal{S}_{11}&=(\Cos)^{-2N}\sin^2\vartheta\cos^6\vartheta\left[\tan^2\vartheta
					-\e^{-\ii h_{v_1} t}\Prod_{\tilde v \neq v_2}w_\vartheta( J_{v_1\tilde v}, t )
					-\e^{-\ii h_{v_2} t}\Prod_{\tilde v \neq v_1}w_\vartheta( J_{v_2\tilde v}, t )
					\right.\\
					&
					\left.
					+\tan^2\vartheta \,\e^{-\ii (h_{v_1}+h_{v_2} + J_{v_1v_2}) t}
					\Prod_{\tilde v \neq v_1,v_2}w_\vartheta( J_{v_1\tilde v} + J_{v_2\tilde v}, t )
					\right]
    	\end{aligned}
    \end{equation}

    \begin{equation}
    	\begin{aligned}
    	\mathcal{S}_{10}&=(\Cos)^{-2N}\sin^2\vartheta\cos^6\vartheta\left[\tan\vartheta
					-\e^{\ii h_{v_2} t}\Prod_{\tilde v \neq v_1}w_\vartheta( -J_{v_2\tilde v}, t )\right.\\
					&-\tan^{-1}\vartheta\e^{-\ii h_{v_1} t}\Prod_{\tilde v \neq v_2}w_\vartheta( J_{v_1\tilde v}, t )\\
					&-\tan^{-2}\vartheta\e^{-\ii h_{v_1} t}\Prod_{\tilde v \neq v_2}w_\vartheta( J_{v_1\tilde v}, t )\\
					&
					\left.
					+\tan^2\vartheta \,\e^{-\ii (h_{v_1}+h_{v_2} + J_{v_1v_2}) t}
					\Prod_{\tilde v \neq v_1,v_2}w_\vartheta( J_{v_1\tilde v} + J_{v_2\tilde v}, t )
					\right]
    	\end{aligned}
    \end{equation}

\begin{equation}
		\begin{aligned}
\langle n_{v_1}n_{v_2}(t)\rangle_\Graph =4\sin^4\vartheta\cos^4\vartheta\,\,\Re &\left\{1
		-\e^{\ii h_{v_1} t}\Prod_{\tilde v\neq v_2}\left({\cos^2\vartheta+\sin^2\vartheta\;\e^{\ii J_{v_1\tilde v}t}}\right)
		-\e^{\ii h_{v_2} t}\Prod_{\tilde v\neq v_1}\left({\cos^2\vartheta+\sin^2\vartheta\;\e^{\ii J_{v_2\tilde v}t}}\right)\right.\\
		&+\frac{1}{2}\e^{\ii (h_{v_1}+h_{v_2} +J_{v_1v_2}) t}\Prod_{\tilde v\neq v_1,v_2}\left({\cos^2\vartheta+\sin^2\vartheta\;\e^{\ii (J_{v_1\tilde v}+J_{v_2\tilde v})t}}\right)\\
		&\left.+\frac{1}{2}\e^{\ii (h_{v_1}-h_{v_2}) t}\Prod_{\tilde v\neq v_1,v_2}\left({\cos^2\vartheta+\sin^2\vartheta\;\e^{\ii (J_{v_1\tilde v}-J_{v_2\tilde v})t}}\right)
\right\}
		\end{aligned}
	\end{equation}

  \textit{$\langle n_{v_1}n_{v_2}(t)\rangle_\Graph - \langle n_{v_1}(t)\rangle_\Graph\langle n_{v_2}(t)\rangle_\Graph$}
  
  In order to emphasise the structure of the expression, we introduce
  
  \begin{equation}
  	w_{v\tilde v}(\vartheta,t) = \left({\cos^2\vartheta+\sin^2\vartheta\;\e^{\ii J_{v\tilde v}t}}\right)
  \end{equation}
  
  and 
  
  \begin{equation}
  	\varrho_{\vartheta,v}(t) = \e^{\ii h_{v_1} t}\Prod_{\tilde v}\left({\cos^2\vartheta+\sin^2\vartheta\;\e^{\ii J_{v\tilde v}t}}\right) = \e^{\ii h_{v_1} t}\Prod_{\tilde v} w_{v\tilde v}(\vartheta,t),
  \end{equation}
  
  so that
  
  \begin{equation}
\langle n_{v}(t)\rangle_\Graph =2\sin^2\vartheta\cos^2\vartheta\,\,\Re \left\{1-\varrho_{\vartheta,v}(t)\right\} = 2\sin^2\vartheta\cos^2\vartheta\left[1-\frac{1}{2}\left(\varrho_{\vartheta,v}(t)+\varrho^*_{\vartheta,v}(t)\right)\right],
\end{equation}

and 
  
  \begin{equation}
\langle n_{v_1}n_{v_2}(t)\rangle_\Graph =4\sin^4\vartheta\cos^4\vartheta\,\,\Re \left\{1
												-\left[\varrho_{\vartheta,v_1}(t)+\varrho_{\vartheta,v_2}(t)\right]
												+\frac{1}{2}\left[\varrho_{\vartheta,v_1}(t)\varrho_{\vartheta,v_2}(t)+\varrho_{\vartheta,v_1}(t)\varrho^*_{\vartheta,v_2}(t)\right]\right\}.
\end{equation}

Similarly,
  \begin{equation}
  \begin{aligned}
\langle n_{v_1}n_{v_2}(t)\rangle_\Graph =4\sin^4\vartheta\cos^4\vartheta\,\,\Re 
		&\left\{1-\frac{\varrho_{\vartheta,v_1}(t)+\varrho_{\vartheta,v_2}(t)}{w_{v_1v_2}(\vartheta,t)}\right.\\
		&+\frac{1}{2}\e^{\ii J_{v_1v_2}t}\frac{\varrho_{\vartheta,v_1}(t)\varrho_{\vartheta,v_2}(t)}
				{\Prod_{\tilde v \in \mathcal{N}_{12}\cup \{v_1,v_2\}}  \cos^2\vartheta+\sin^2\vartheta\;\e^{\ii (J_{v_1\tilde v}+J_{v_2\tilde v})t}}\\
		&\left.+\frac{1}{2}\frac{\varrho_{\vartheta,v_1}(t)\varrho_{\vartheta,v_2}^*(t)}
				{\Prod_{\tilde v \in \mathcal{N}_{12}\cup \{v_1,v_2\}}  \cos^2\vartheta+\sin^2\vartheta\;\e^{\ii (J_{v_1\tilde v}-J_{v_2\tilde v})t}}
\right\}.
\end{aligned}
\end{equation}

Introducing $w^{\pm}_{v1v_2}(\vartheta,t)=\Prod_{\tilde v \in \mathcal{N}_{12}\cup \{v_1,v_2\}}\cos^2\vartheta+\sin^2\vartheta\;\e^{\ii (J_{v_1\tilde v}\pm J_{v_2\tilde v})t}$, the correlation between densities at $v_1$ and $v_2$ can then be expressed as

\begin{tcolorbox}
\begin{align}
\langle n_{v_1}n_{v_2}(t)\rangle_\Graph -\langle n_{v_1}(t)\rangle_\Graph\langle n_{v_2}(t)\rangle_\Graph=4\sin^4\vartheta\cos^4\vartheta\,\,\Re 
		&\left\{\left[\varrho_{\vartheta,v_1}(t)+\varrho_{\vartheta,v_2}(t)\right]\left[1-w_{v_1v_2}(\vartheta,t)^{-1}\right]\right. \nonumber \\
		&+\frac{1}{2}\left[1-\e^{\ii J_{v_1v_2}t} w^+_{v_1v_2}(\vartheta,t)^{-1}\right]
		\varrho_{\vartheta,v_1}(t)\varrho_{\vartheta,v_2}(t) \nonumber \\
		&\left.+\frac{1}{2}\left[1- w^-_{v_1v_2}(\vartheta,t)^{-1}\right]
		    \varrho_{\vartheta,v_1}(t)\varrho_{\vartheta,v_2}^*(t)
\right\}.
\label{eq:generic_corr}
\end{align}
\end{tcolorbox}

Since we consider only the uniform Ising model for this proof, we have $h_v = 1 \; \forall v \in \mathcal{V}$ and $J_{v_1,v_2} = 1 $ if $(v_1,v_2) \in \mathcal{E}$ and $0$ if not. This allows us to rewrite the elements recovered in the generic expression derived in equation \ref{eq:generic_corr}. furthermore, if we write $a = (\cos^2\vartheta + \sin^2\vartheta e^{\ii t})$ and $b = (\cos^2\vartheta + \sin^2\vartheta e^{2\ii t})$ we can rewrite the previous quantities in terms of SRGs invariants $(\nu,k,\lambda,\mu)$ as :

\begin{equation}
\varrho_{\vartheta,v}(t) = \e^{\ii t}a^{k}
\end{equation} 

\begin{equation}
  w^{+}_{v_1,v_2}(\vartheta,t) = \left\{
  \begin{array}{@{}ll@{}}
    a^2 . b^{\lambda}, & \text{if}\ (v_1,v_2) \in \edges \\
     b^{\mu} & \text{otherwise}
  \end{array}\right.
\end{equation}

\begin{equation}
  w^{-}_{v_1,v_2}(\vartheta,t) = \left\{
  \begin{array}{@{}ll@{}}
    |a|^2, & \text{if}\ (v_1,v_2) \in \edges \\
     1 & \text{otherwise}
  \end{array}\right.
\end{equation} 

this allows us to write the correlation matrix $C$ as : 
\begin{equation}
C = c_{k,\lambda} A + c_{k,\mu} (J-A) 
\end{equation} 
with   $c_{k,\lambda}$ \&  $c_{k,\mu} \in \mathbb{R}$ that only depend on the invariants, $A$ the adjacency matrix of the graph, and $J$ a matrix where all the entries are ones. It is then easy to verify that using this formula, invariant functions (ex :  the distance introduced in section \ref{sect:theory_srg}) are not enough to distinguish non isomorphic SRGs from the same family, neither can the GD-WL which is in this case equivalent to the 1-WL, and we know that the latter is not powerful enough to distinguish SRGs \textit{c.f} proposition \ref{pr:3WL_SRG}. 

\section{Experiments}
\label{sec:experiments}

\subsection{Experiments on quantum random walk}
\label{app:qrw_experiment}

In the same way as \citep{ma2023graph}, we perform the experiments on the standard train/val/test splits. For each dataset, we perform 4 runs with the seeds 0, 1, 2, 3 and display the average of the scores and the standard deviation. 

We do not perform an extensive hyperparameter search, and we only compute ourselves the GRIT model. We take the same hyperparameters as \citep{ma2023graph} that we remind in table \ref{tab:hyperparams_grit}.

\begin{table}[h!]
    \centering
    \caption{Hyperparameters for GRIT model five datasets from BenchmarkingGNNs \citep{dwivedi2020benchmarking},  ZINC-full~\citep{irwin2012zinc} and \citep{hu2021ogb}
    }
    \label{tab:hyperparams_grit}
    {\scriptsize
\begin{tabular}{lcccccc}
\toprule
Hyperparameter & ZINC/ZINC-full & MNIST & CIFAR10 & PATTERN & CLUSTER & PCQM4Mv2 \\
\midrule
\# Transformer Layers & 10 & 3 & 3 & 10 & 16 & 16\\
Hidden dim & 64 & 52 & 52 & 64 & 48 & 256 \\
\# Heads & 8 & 4 & 4 & 8 & 8 & 8 \\
Dropout & 0 & 0 & 0 & 0 & $0.01$ & $0.1$ \\
Attention dropout & $0.2$ & $0.5$ & $0.5$ & $0.2$ & $0.5$ & $0.1$ \\
Graph pooling & sum & mean & mean & $-$ & $-$ & mean \\
\midrule
PE dim (RW-steps) & 21 & 18 & 18 & 21 & 32 & 16 \\
PE encoder & linear & linear & linear & linear & linear & linear \\
\midrule
QPE dim (1CQRW steps) & 20 & 18 & 18 & 20 & 32 & 16 \\
Max duration & $\pi$ & $\pi$ & $\pi$ & $\pi$ & $\pi$ & $\pi$\\
Min duration & $0.1$ & $0.1$  & $0.1$  & $0.1$  & $0.1$  & $0.1$ \\
Initial distribution & local & local & local & local & local & local\\
\midrule
QPE dim (2QiRW steps) & 20 & 18 & 18 & 20 & 32 & 16 \\
Initial distribution & adjacency & adjacency & adjacency & adjacency & adjacency & adjacency\\
\midrule
Batch size & 32/256 & 16 & 16 & 32 & 16 & 256\\
Learning Rate & $0.001$ & $0.001$ & $0.001$ & $0.0005$ & $0.0005$ & 0.0002 \\
\# Epochs & 2000 & 200 & 200 & 100 & 100 & 150 \\
\# Warmup epochs & 50 & 5 & 5 & 5 & 5 & 10 \\
Weight decay & $1 \mathrm{e}-5$ & $1 \mathrm{e}-5$ & $1 \mathrm{e}-5$ & $1 \mathrm{e}-5$ & $1 \mathrm{e}-5$ & 0\\
\midrule
\# Parameters GRIT & 473,473 & 102,138 & 99486 & 477,953 & 432,206 & 11.8M \\
\# Parameters 2QiRW GRIT & 476,033 & 104,010 & 101,358 & 480,513 & 434,742 & 11.8M\\
\bottomrule
\end{tabular}
}
\end{table}

\begin{table}[h!]
    \centering
    \caption{Hyperparameters for non transformer base models large scale datasets ,  ZINC-full~\citep{irwin2012zinc} and PCQM4Mv2 \citep{hu2021ogb}.  Each entry has to be read as the values for ZINC-full/PCQM4Mv2, when there is a single entry, the value is the same for both datasets. * : same as the first column. - : non applicable.
    }
    \label{tab:hyperparams_large}
    {\scriptsize
\begin{tabular}{lcccccc}
\toprule
Hyperparameter & GINE & GINE-big & GatedGCN & GatedGCN-big & &\\
\midrule
\# Layers & 5/3 & * & * & * \\
Hidden dim & 128 & 256 & 128 & 256 & \\
Dropout & 0 & * & * & *  \\
Aggregation & mean & * & * & *  \\
\# Layers MLP postprocessing & 3 & * & * & *&&\\
PE encoder & linear & *&*&* \\
\midrule
PE dim (RRWP) & 21 & 40 & 21 & 40 \\
\midrule
PE dim (LE) & 32 & * &* &* \\
\midrule
PE dim (Q) & 21 & 40 & 20 &40 \\
Initial distribution & adjacency&*&*&*\\
\midrule
PE dim RRWP(RRWP+Q) & 20 & * &* &* \\
PE dim Q(RRWP+Q) & 20 & * &* &* \\
\midrule
Batch size & 256 & * & * & * \\
Learning Rate & $0.001/0.0002$ & * & * &* \\
\# Epochs & 2000/150 & * & * & * \\
\# Warmup epochs & 50/10 & * & * & * \\
Weight decay & $1 \mathrm{e}-5/0$ & * & * &*\\
\midrule
\bottomrule
\end{tabular}
}
\end{table}

\begin{table}[h!]
\centering
    \caption{Test performance on ZINC-full and PCQM4MV2.
    }
    {\small
    \begin{tabular}{clcc}
    \toprule
       \textbf{Method} & \textbf{Model}  & 
       \textbf{ZINC-full} (MAE $\downarrow$) &
       \textbf{PCQM4MV2} (MAE $\downarrow$) 
       \\ \midrule
        \multirow{4}{*}{GatedGCN}
       & LE & $.033 \pm .001$ & $.1056$ \\
       & RRWP & $.026 \pm .003$ & $.1045$ \\
       & Q & $.031 \pm .002$ & $.1079$\\
       & RRWP+Q & $.026 \pm .001$ & $.1052$\\
       \midrule
       \multirow{4}{*}{GatedGCN-big}
       & LE & $.033 \pm .0008$ & $.1016$ \\
       & RRWP & $.025 \pm .0017$ & $.1005$ \\
       & Q & $.025 \pm .0023$ & $.1035$\\
       & RRWP+Q & $.022 \pm .0017$ & $.0999$\\
       \midrule
       \multirow{4}{*}{GINE}
       & LE & $.035 \pm .002$ & $.1155$ \\
       & RRWP & $.029 \pm .003$ & $.114 $\\
       & Q & $.027 \pm .0005$ &$.1149$\\
       & RRWP+Q & $.029 \pm .003$& $.1124$\\
       \midrule 
       \multirow{4}{*}{GINE-big}
       & LE & $.036 \pm .0022$ & $.1063$ \\
       & RRWP & $.029 \pm .003$ & $.1041$ \\
       & Q & $.024 \pm .002$ & $.1054$\\
       & RRWP+Q & $.027 \pm .0025$ & $.1048$\\
       \midrule
       \multirow{2}{*}{GRIT}
       & RRWP & $0.025 \pm 0.002$ & $.0842$\\
       & RRWP+Q & $0.023 \pm 0.002$ & $.0838$\\
       \bottomrule
    \end{tabular}
    }
    \label{tab:large_datasets_numbers}
\end{table}

\subsection{Experiments on synthetic datasets}
\label{app:syth_experiment}

 We train  the GCN model for 200 epochs using the Adam optimizer, 0.001 learning rate, no weight decay. We split randomly the dataset on train/validation/test with a proportion 0.8/0.1/0.1, and we measure the test accuracy of the model having the highest validation accuracy. For the other models we use the same hyperparameters as in table \ref{tab:hyperparams_large}, but with hidden dimensions of 32 for normal models and 64 for big models. We also use a dimension 20 for all positional encodings, and initialize with uniform node and edge features full of 1s.

\section{Supplementary information about the datasets}
\label{app:datasets}

\subsection{Datasets used in experiments on quantum random walks}

The datasets used for benchmarking the use of quantum random walks encodings are standard in the GNN community. The first five are from \citep{dwivedi2020benchmarking}, the last one is from \citep{hu2021ogb}. We reproduce the table of statistics \ref{tab:dataset_qrw} taken from \citep{ma2023graph}, and we also refer the reader to \citep{rampavsek2022recipe} for more information about the datasets.

\begin{table}[h!]
    \centering
    \caption{Overview of the graph learning datasets involved in this work \citep{dwivedi2020benchmarking}, \citep{irwin2012zinc}, \citep{hu2021ogb} .}
    \small
    \setlength{\tabcolsep}{1.6pt}
    {\scriptsize
    \begin{tabular}{l|ccccccc}
    \toprule
       \textbf{Dataset} & \textbf{\# Graphs} & \textbf{Avg. \# nodes} & \textbf{Avg. \# edges}  &  \textbf{Directed} 
 & \textbf{Prediction level} & \textbf{Prediction task} & \textbf{Metric}\\
 \midrule
        ZINC(-full) & 12,000 (250,000) & 23.2 & 24.9 & No &  graph & regression & Mean Abs. Error \\
        MNIST &  70,000  &70.6  & 564.5 & Yes  & graph & 10-class classif. &  Accuracy \\
        CIFAR10 & 60,000 & 117.6 & 941.1 & Yes  & graph & 10-class classif. & Accuracy \\
        PATTERN & 14,000 & 118.9 & 3,039.3  & No & inductive node & binary classif. & Weighted Accuracy \\
        CLUSTER & 12,000 & 117.2 & 2,150.9 &  No & inductive node &  6-class classif. & Accuracy \\ 
        \midrule
        PCQM4Mv2 & 3,746,620 & 14.1 & 14.6 & No & graph & regression & Mean Abs. Error \\
        \bottomrule
    \end{tabular}
    }
    \label{tab:dataset_qrw}
\end{table}

\subsection{Synthetic Dataset}
\label{app:synthetic_data}

\begin{figure}[h!]

         \includegraphics[width=\textwidth]{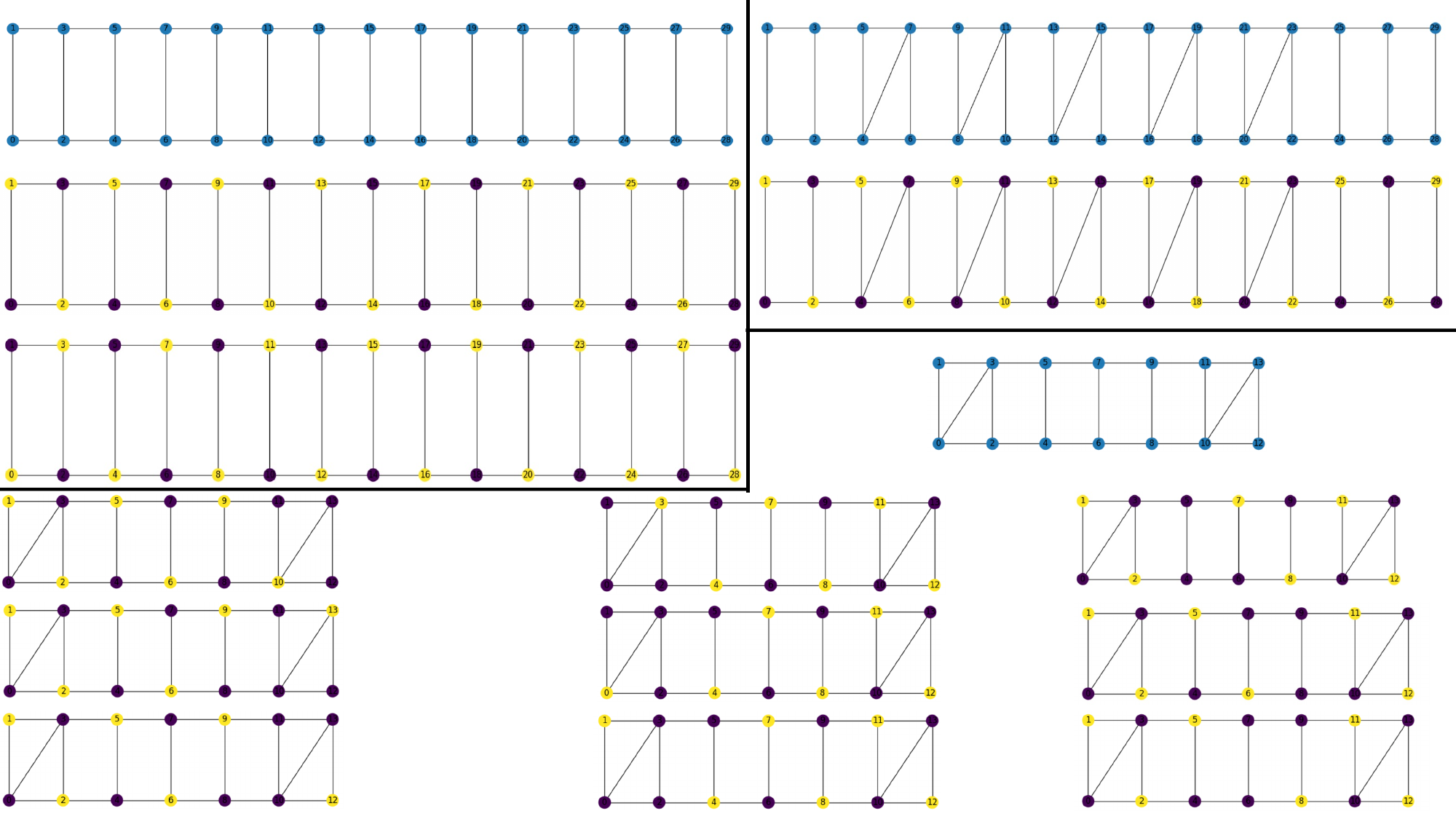}
         \caption{The base subgraphs (type 0, type 1, type 2) and their possible ground state. Top left: type 0 graph of length 15, 2 possible ground states. Top right: type 1 graph of length 15, 1 possible ground state. Bottom : type 2 graph of length 9, 9 possible ground states.}
         \label{fig:ladder_subgraph}

\end{figure}

In this subsection, we explain how to construct our artificial dataset. Our building blocks are 3 types of graphs, called types 0, 1, 2. Each type is composed of one ladder graphs with crossings inserted at different places. All crossings are in the same fixed arbitrary direction. Type 0 graphs are plain ladder graphs and their Ising hamiltonian has two ground states. Type 1 graphs are type 0 graphs with crossings separated with an odd number of nodes. The crossings are located such that they have one possible Ising ground state which is one of the ground states of the type 0 associated graph. The crossings will effectively select one of the two possible ground states. Type 2 graphs are ladder graphs of odd length with crossings at the beginning and the end. An illustration of the types of graphs is provided figure \ref{fig:ladder_subgraph}. 

\begin{figure}[h!]

         \includegraphics[width=.6\textwidth]{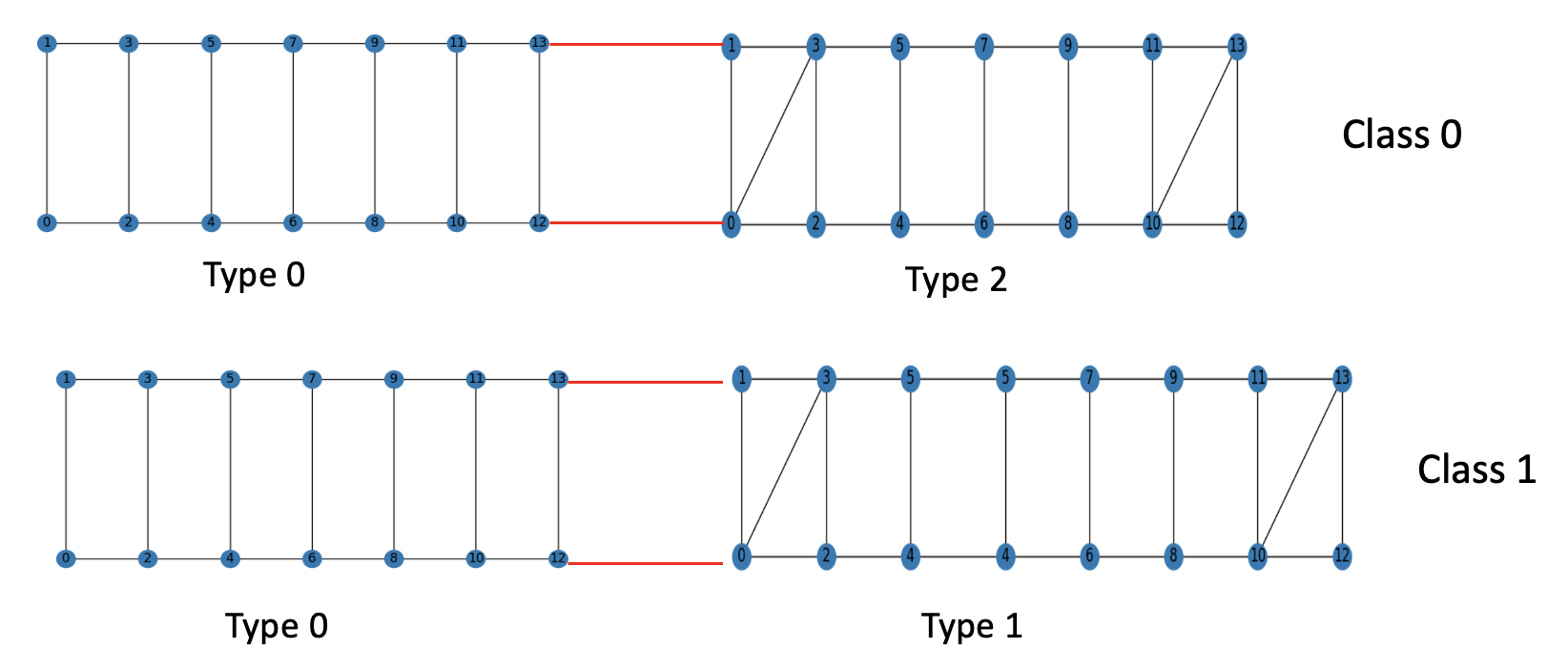}
         \includegraphics[width=.4\textwidth]{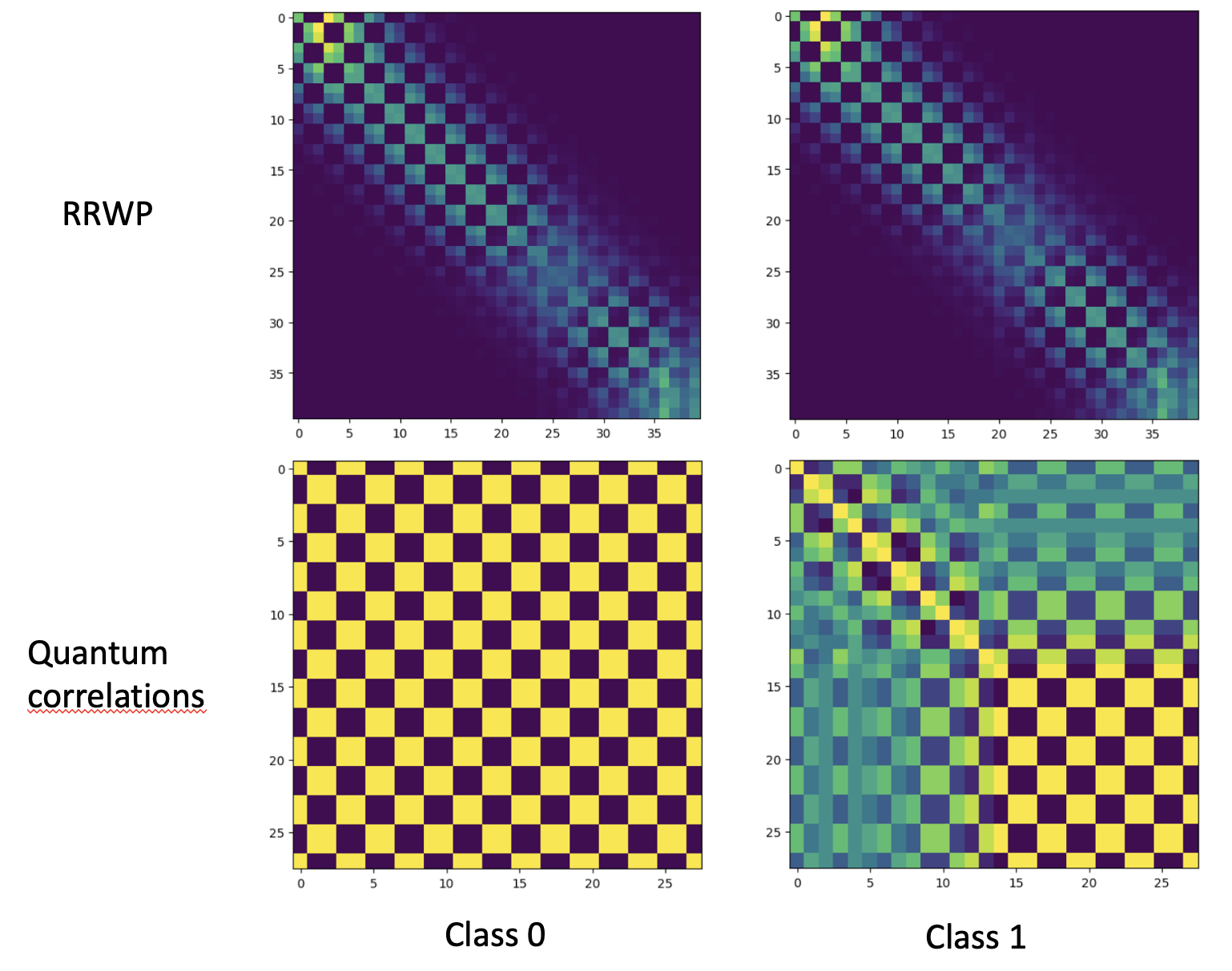}
    \caption{Left: construction of our arificial dataset. Right: RRWP and quantum features for each class of the dataset on a 40 nodes graph.}
\label{fig:ladder_dataset}
\end{figure}

We construct a graph given two graphs of same length but different types concatenated to each other. The first class is determined by graphs of type 0 and type 1 concatenated, and the second class is composed of graphs of type 0 concatenated to graphs of type 1.  The concatenation is made by adding edges to continue the ladder, the process is illustrated figure \ref{fig:ladder_dataset}. The ground state of the total graph is included in a union of the groundstates of the subgraph, so it can be efficiently computed. 
The length of graphs are taken between 100 and 400, our dataset consists of 400 graphs per class, so 800 in total.

In figure \ref{fig:ladder_dataset} we see that the RRWP features are very similar for the two classes whereas the correlations on the ground state are very different.



\end{document}